%% file: aaai.tex
\documentclass[letterpaper]{article} 
\usepackage{aaai2026}  
\usepackage{times}  
\usepackage{helvet}  
\usepackage{courier}  
\usepackage[hyphens]{url}  
\usepackage{graphicx} 
\usepackage{natbib}  
\usepackage{caption} 
\usepackage{algorithm}
\usepackage{algorithmic}
\usepackage{amsmath} 
\usepackage{amssymb} 
\usepackage{amsthm}
\usepackage[table]{xcolor}  
\usepackage{multirow}

\usepackage{booktabs}
\usepackage{subcaption}
\newcolumntype{P}[1]{>{\raggedright\arraybackslash}p{#1}}  
\usepackage{newfloat}
\usepackage{listings}
\DeclareCaptionStyle{ruled}{labelfont=normalfont,labelsep=colon,strut=off} 
\floatstyle{ruled}
\newfloat{listing}{tb}{lst}{}
\floatname{listing}{Listing}
\graphicspath{{fig/}}

\title{Ghost in the Transformer: Detecting Model Reuse with Invariant Spectral Signatures}
\author {
    Suqing Wang\textsuperscript{\rm 1}\equalcontrib,
    Ziyang Ma\textsuperscript{\rm 2}\equalcontrib,
    Li Xinyi\textsuperscript{\rm 2},
    Zuchao Li\textsuperscript{\rm 1}\thanks{Corresponding author.}
}
\affiliations {
    \textsuperscript{\rm 1}School of Artificial Intelligence, Wuhan University\\
    \textsuperscript{\rm 2}School of Computer Science, Wuhan University\\
    {\{wangsuqing, maziyang, xyli-lucia, zcli-charlie\}@whu.edu.cn}
}

\begin{document}

\maketitle

\begin{abstract}
Large Language Models (LLMs) are widely adopted, but their high training cost leads many developers to fine-tune existing open-source models. While most adhere to open-source licenses, some falsely claim original training despite clear derivation from public models, raising pressing concerns about intellectual property protection and the need to verify model provenance.
In this paper, we propose GhostSpec, a lightweight yet effective method for verifying LLM lineage without access to training data or modification of model behavior. Our approach constructs compact and robust fingerprints by applying singular value decomposition (SVD) to invariant products of internal attention weight matrices. Unlike watermarking or output-based methods, GhostSpec is fully data-free, non-invasive, and computationally efficient. Extensive experiments show it is robust to fine-tuning, pruning, expansion, and adversarial transformations, reliably tracing lineage with minimal overhead. By offering a practical solution for model verification, our method contributes to intellectual property protection and fosters a transparent, trustworthy LLM ecosystem. 
\end{abstract}

\begin{links}
    \link{Code}{https://github.com/DX0369/GhostSpec}
    \link{Extended version}{http://arxiv.org/abs/2511.06390}
\end{links}

\section{Introduction}

LLMs have quickly become essential for various applications in research and industry~\citep{achiam2023gpt, yang2025qwen3, wang2025parameters, zhang2025segment, poon2025online}. Due to the high cost of training LLMs from scratch~\citep{workshop2022bloom}, many developers modify open-source LLMs via fine-tuning, continued pre-training, merging, and compression~\citep{yang2024model, zhu2024survey, tang2025covipal, yang2025xquant, li2023enhancing, hu2025songsong}. While most developers comply with open-source licenses, there have been instances of falsely claiming to have trained models “from scratch” when they are in fact repackaged or fine-tuned versions of public models (e.g., Llama3-V and MiniCPM-Llama3-V 2.5)~\citep{yao2024minicpm}. It is crucial to distinguish such intellectual property violations, which often break attribution requirements, from legitimate, licensed fine-tuning. This raises concerns about plagiarism and intellectual property violations, emphasizing the need for tools to verify model lineage.

To address these concerns, researchers have proposed various model identification methods~\citep{sun2023deep}, which can be broadly classified into black-box and white-box approaches. Black-box methods identify models without accessing their weights, using techniques like behavioral fingerprinting and watermarking. However, these approaches are often sensitive to randomness, adversarial changes, or require intrusive pipeline modifications. In contrast, white-box methods leverage internal parameters. While representation-based techniques analyze hidden states or gradients, they depend on data access and are computationally expensive. Direct weight comparisons are simpler but fragile under fine-tuning or pruning.

We propose GhostSpec, a simple, data-free, and robust white-box method for verifying LLM lineage. GhostSpec is, by design, a white-box method targeting the open-weight model ecosystem where weights are accessible. Our key insight is that the spectral structure of weight matrices encodes intrinsic information about a model’s origin, remaining stable under various modifications. Specifically, we apply singular value decomposition (SVD) to internal matrix products within the attention mechanism, including the query-key and value-output weight products. From the resulting spectra, we select the most significant singular values based on effective rank to compute similarity. To handle architectural variations in depth, such as those resulting from layer pruning or expansion, we develop the Penalty-based Optimal Spectral Alignment (POSA) algorithm that finds the best layer-wise correspondence between models. This yields a quantitative similarity score robust to differences in depth and architectural variations. Unlike black-box or representation-based white-box methods, GhostSpec is data-independent, requires no model modification, and has minimal computational cost.

\begin{figure*}[t]
    \centering
    \includegraphics[width=1\textwidth]{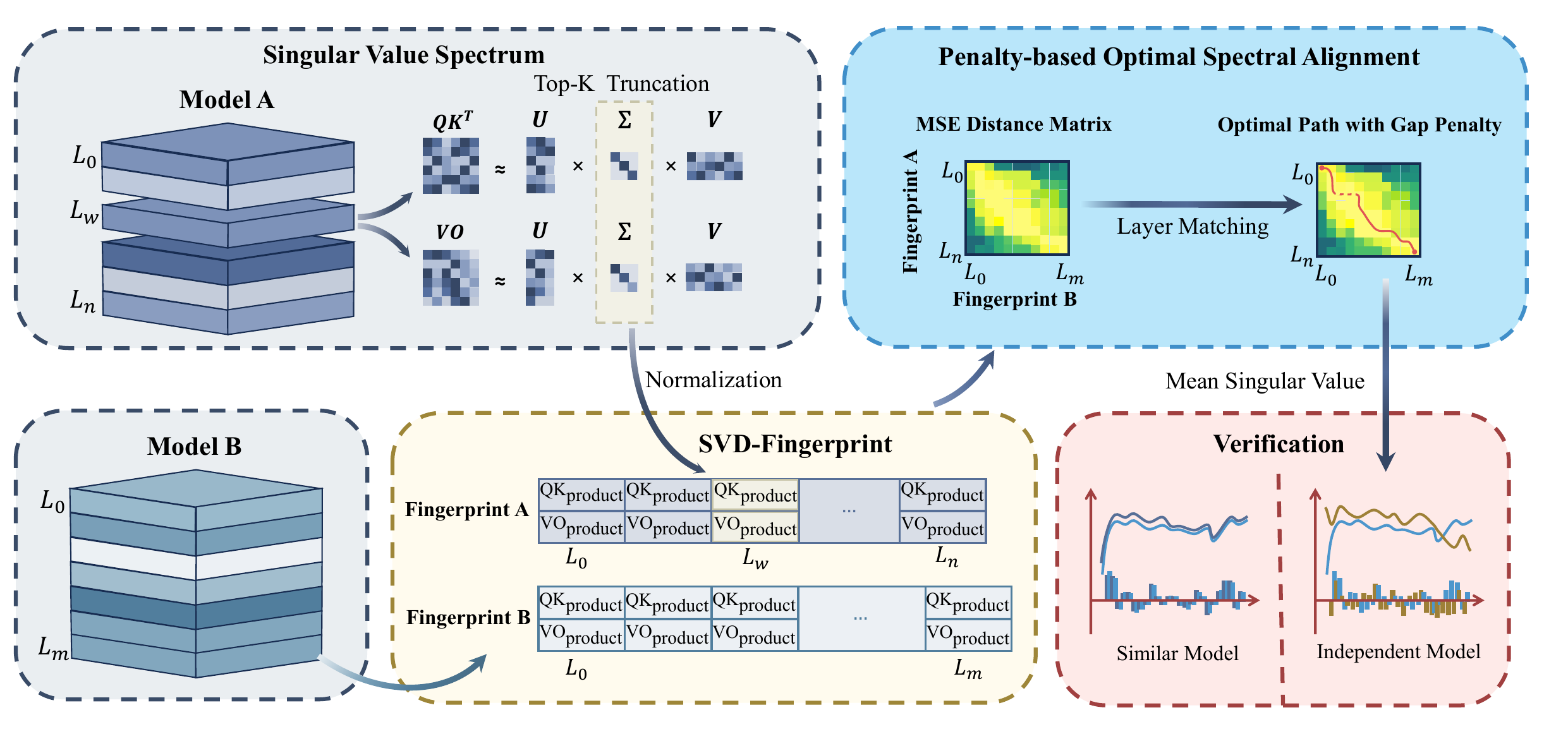}
    \caption{GhostSpec extracts singular value spectra from each layer’s attention products to form spectral fingerprints. A pairwise MSE distance matrix is computed, and a penalty-based alignment algorithm matches layers across models of different depths. The final similarity score distinguishes between related and independently trained models.
    }
    \label{fig:main_figure}
\end{figure*}

\noindent
\textbf{Our contributions are summarized as follows:}
\begin{itemize}
    \item We propose GhostSpec, a lightweight white-box method for verifying LLM lineage from model weights, requiring no training data or architectural changes, offering a practical solution for provenance verification and IP protection in open-source LLMs.

    \item We introduce spectral fingerprints based on invariant matrix products within the attention mechanism, robust to scaling and permutation transformations, and develop the POSA algorithm for comparing models with varying depths and architectures.

    \item We demonstrate through experiments that GhostSpec reliably distinguishes derivative models from independently trained ones, even under challenging modifications.
\end{itemize}

\section{Related Work}

Efforts to verify model lineage can be broadly classified into black-box and white-box approaches, depending on whether internal model access is required.

\subsection{Black-Box Identification}
Black-box methods operate without access to model weights and are suitable for closed-source, API-only models. They include behavioral fingerprinting and watermarking.

\subsubsection{Behavioral Fingerprinting.}
These passive methods identify model signatures from natural outputs. Approaches analyze stylistic or statistical patterns in generated text, or use crafted prompts to probe responses~\citep{pasquini2024llmmap, mcgovern2024your, sam2025predicting}. Some rely on output logits or top-k probabilities to define a unique model space~\citep{yang2024fingerprint}. However, such methods are sensitive to decoding randomness and vulnerable to adversarial paraphrasing.

\subsubsection{Watermarking.}
Watermarking embeds a detectable signal in model outputs, either via instruction tuning~\citep{xu2024instructional} or token-level perturbation~\citep{kirchenbauer2023watermark, nagatsuka2025nested}. These signals can be verified statistically, but require the model creator’s cooperation and can be invalidated by output editing or algorithm exposure.

\subsection{White-Box Identification}

White-box approaches leverage internal model weights or activations, typically by computing similarity between weights, intermediate representations, or gradients.

\subsubsection{Representation-based Fingerprinting.}
These methods analyze hidden representations and generally require input data. Techniques such as CKA similarity or gradient statistics have been used to reveal shared training origins~\citep{zhang2024reef, wu2025gradient, liang2025origin}. While effective, these methods are computationally intensive, data-dependent, and may raise concerns regarding potential correlations with training data.

\subsubsection{Weight-based Fingerprinting.}
This line of work focuses on static, data-free analysis of model weights. Prior methods include visualizing invariant structural features~\citep{zeng2024huref} or analyzing layer-wise statistics~\citep{yoon2025intrinsic}. Our proposed GhostSpec method falls into this category, capturing deeper structural information by leveraging the full singular value spectrum of invariant matrix products. The POSA algorithm is further introduced to robustly trace model ancestry under various transformations.

\begin{figure}[t]
    \centering
    \includegraphics[width=0.47\textwidth]{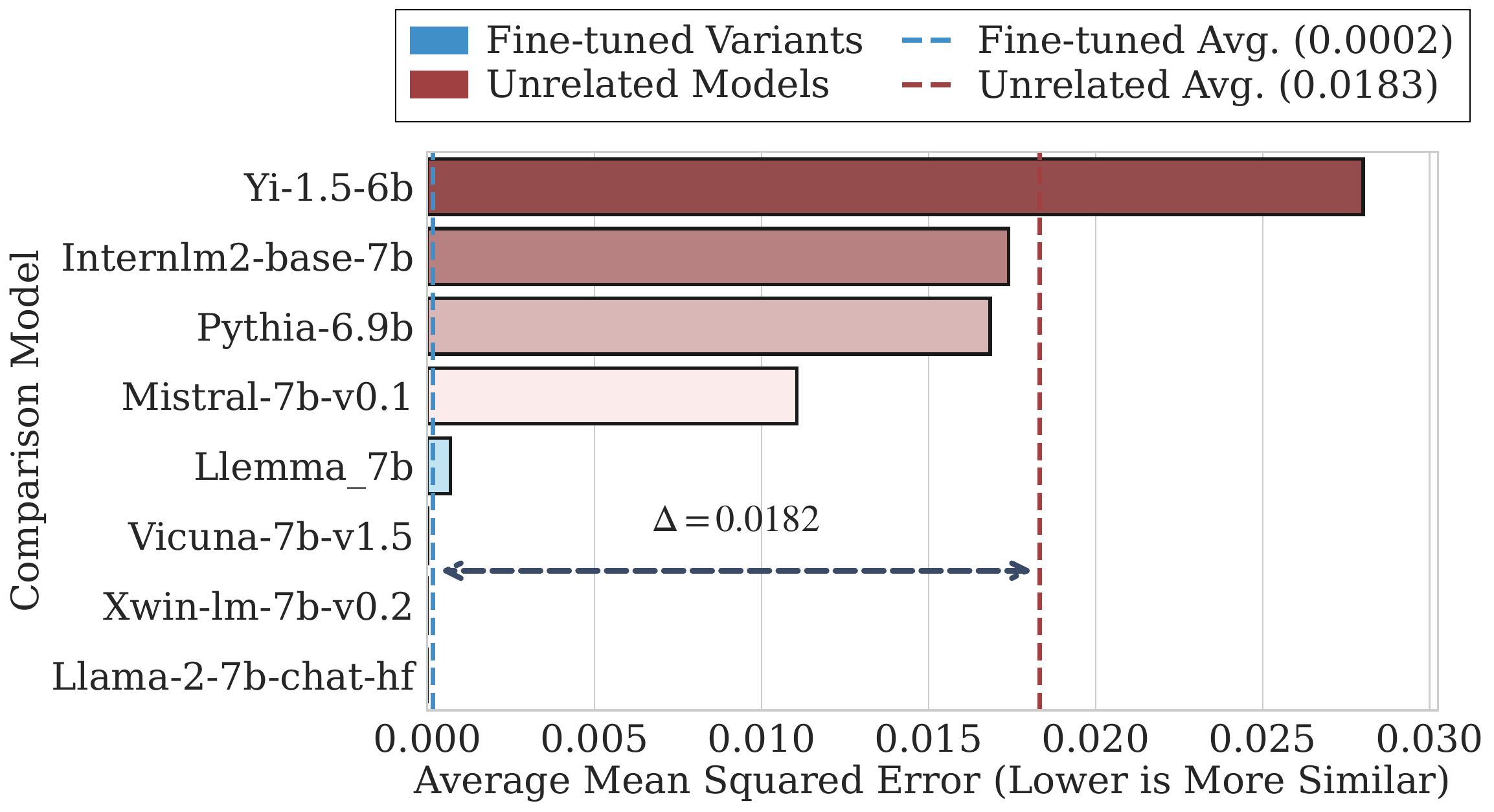}
    \caption{Average MSE of normalized singular values from Q/K/V/O projections. The spectral distance from Llama-2-7b to its fine-tuned variants (blue) is negligible, while the distance to unrelated models (red) is large, confirming the fingerprint's robustness against fine-tuning.}
    \label{fig:preliminaries_singular_mse}
\end{figure}

\begin{figure}[t]
    \centering
    \includegraphics[width=\columnwidth]{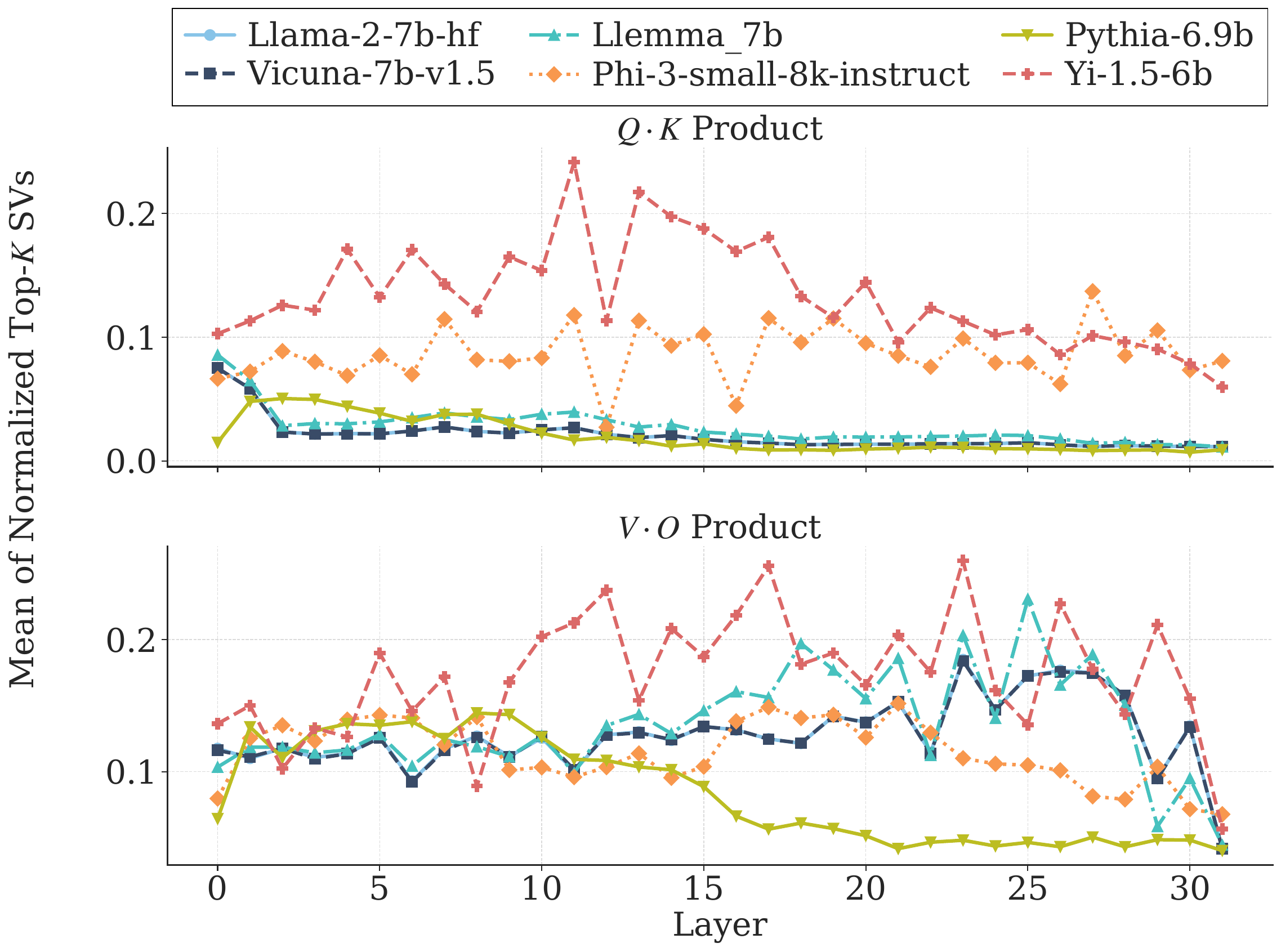}
    \caption{The layer-wise trend of the mean of normalized singular values for various models. Models with a shared lineage (e.g., Llama-2-7b-hf and its variants) exhibit highly correlated trends, while unrelated models show divergent patterns.}
    \label{fig:layers_trend_norm}
\end{figure}

\section{Preliminaries}
\citet{staats2024small} use Random Matrix Theory (RMT) to analyze the singular value spectrum of pre-trained language models. Their analysis reveals deviations from the Marchenko--Pastur distribution, especially in large singular values. These outliers correspond to dominant directions in the weight space and are strongly associated with the model's learned representations. Removing them significantly increases perplexity, showing they form a stable backbone of the model's identity.

Fine-tuning, in contrast to training from scratch, refines rather than rebuilds the model’s internal structure. \citet{staats2024small} demonstrates that fine-tuning primarily affects directions associated with small singular values. A key asymmetry is observed: pruning small singular values after fine-tuning leads to significantly greater performance degradation than pruning them before fine-tuning. This indicates that fine-tuning updates are concentrated in low-magnitude spectral components.

Together, these findings suggest spectral stability: large singular values, encoding foundational pre-trained knowledge, remain stable throughout adaptation and anchor the model’s global behavior. In contrast, fine-tuning introduces targeted, low-rank modifications reflected in small singular values. We hypothesize that the large singular value spectrum of attention matrices serves as a robust, fine-tuning-invariant fingerprint.

\subsection{Experimental Design}
To empirically test this hypothesis, we designed an experiment centered on Llama-2-7b-hf. We compared this primary model against two distinct groups:
\begin{itemize}
    \item \textbf{Fine-tuned Variants:} A group of known direct descendants of Llama-2-7b~\citep{touvron2023llama}, including Llama-2-7b-chat-hf, Vicuna-7b-v1.5, Llemma\_7b~\citep{azerbayev2023llemma}, and Xwin-LM-7B-V0.2~\citep{xwin-lm}.
    \item \textbf{Unrelated Models:} A control group of architecturally distinct models, including Mistral-7B-v0.1~\citep{DBLP:journals/corr/abs-2310-06825}, Pythia-6.9b~\citep{biderman2023pythia}, Yi-1.5-6B~\citep{young2024yi}, Internlm2-base-7b~\citep{cai2024internlm2}.
\end{itemize}
Our objective is to quantify the spectral distance between the base model and each model in these groups.

\subsection{Quantifying Spectral Similarity}
To quantify the similarity between the singular value spectra of two models $A$ and $B$, we define a compact layer-wise distance metric.

For each projection type $p \in \{q, k, v, o\}$ and each layer $i \in \{1, \dots, L\}$, we extract the singular value vectors $\mathbf{s}_{p,A}^{(i)}$ and $\mathbf{s}_{p,B}^{(i)}$ from the corresponding weight matrices. We first truncate both vectors to their minimum effective rank $r_p^{(i)}$ and then apply min-max normalization to map them into $[0, 1]$. The spectral distance at layer $i$ and projection $p$ is then defined as the Mean Squared Error (MSE) between the normalized, truncated vectors:
\begin{equation}
\begin{split}
    d_p^{(i)}(A, B) = \frac{1}{r_p^{(i)}} \left\lVert \right. & \text{norm}(\text{trunc}(\mathbf{s}_{p,A}^{(i)})) \\
    & \left. - \text{norm}(\text{trunc}(\mathbf{s}_{p,B}^{(i)})) \right\rVert_2^2.
\end{split}
\end{equation}

The overall spectral distance is computed by averaging over all layers and projection types:
\begin{equation}
D(A, B) = \frac{1}{4L} \sum_{p \in \{q, k, v, o\}} \sum_{i=1}^{L} d_p^{(i)}(A, B).
\end{equation}

\subsection{Empirical Validation}
As shown in Figure~\ref{fig:preliminaries_singular_mse}, fine-tuned models have low spectral MSE with their base models, indicating similar structures. In contrast, unrelated models show much higher distances, highlighting structural dissimilarity.

These results confirm that the singular value spectrum is a stable, intrinsic property, mostly preserved through fine-tuning, motivating the design of a robust fingerprinting framework for model lineage verification under transformations.

\definecolor{mygreen}{HTML}{B9E6FA}
\definecolor{myred}{HTML}{FEEAE8}
\definecolor{tableheadercolor}{HTML}{D9DEE7}
\definecolor{primaryheadercolor}{HTML}{B6C2D8}

\newcommand{\colorQueRE}[1]{\ifdim #1pt < 0.18pt \cellcolor{myred}\else\cellcolor{mygreen}\fi #1}
\newcommand{\colorLogits}[1]{\ifdim #1pt < 0.82pt \cellcolor{myred}\else\cellcolor{mygreen}\fi #1}
\newcommand{\colorREEF}[1]{\ifdim #1pt < 0.82pt \cellcolor{myred}\else\cellcolor{mygreen}\fi #1}
\newcommand{\colorPCS}[1]{\ifdim #1pt < 0.12pt \cellcolor{myred}\else\cellcolor{mygreen}\fi #1}
\newcommand{\colorGhostCorr}[1]{\ifdim #1pt < 0.61pt \cellcolor{myred}\else\cellcolor{mygreen}\fi #1}
\newcommand{\colorGhostMSE}[1]{\ifdim #1pt < 0.85pt \cellcolor{myred}\else\cellcolor{mygreen}\fi #1}

\begin{table*}[t]
\centering
\scalebox{0.82}{
\renewcommand{\arraystretch}{1.2}
\begin{tabular}{l|l|cc|cc|cc}
\hline\hline
\multicolumn{8}{c}{\cellcolor{primaryheadercolor}\textbf{Primary Model: Llama-2-7b}} \\
\hline
\multirow{3}{*}{\textbf{Method}} & \multirow{3}{*}{\textbf{Data Dep.}}
& \multicolumn{2}{c|}{\cellcolor{tableheadercolor}\textbf{Model Fine-tuning~($\uparrow$)}}
& \multicolumn{2}{c|}{\cellcolor{tableheadercolor}\textbf{Adversarial Transforms~($\uparrow$)}}
& \multicolumn{2}{c}{\cellcolor{tableheadercolor}\textbf{Unstructured Pruning~($\uparrow$)}} \\
\cline{3-8}
& & \begin{tabular}[c]{@{}c@{}}Vicuna-7b-v1.5\end{tabular}
& Llemma\_7b
& \begin{tabular}[c]{@{}c@{}}Llama-2-7b\\-scaled\end{tabular}
& \begin{tabular}[c]{@{}c@{}}Llama-2-7b\\-permuted\end{tabular}
& \begin{tabular}[c]{@{}c@{}}Pruned-50\%\\-Retrained\end{tabular}
& \begin{tabular}[c]{@{}c@{}}Pruned-70\%\\-Retrained\end{tabular} \\
\hline
\textbf{QueRE} & Data-Aware & \colorQueRE{1.0000} & \colorQueRE{1.0000} & \colorQueRE{1.0000} & \colorQueRE{1.0000} & \colorQueRE{1.0000} & \colorQueRE{1.0000} \\
\textbf{Logits} & Data-Aware & \colorLogits{0.9767} & \colorLogits{0.8400} & \colorLogits{1.0000} & \colorLogits{1.0000} & \colorLogits{0.8567} & \colorLogits{0.8533} \\
\textbf{REEF} & Data-Aware & \colorREEF{0.9992} & \colorREEF{0.9979} & \colorREEF{1.0000} & \colorREEF{1.0000} & \colorREEF{0.9968} & \colorREEF{0.9948} \\
\textbf{PCS} & Data-Free & \colorPCS{0.9986} & \colorPCS{0.5052} & \colorPCS{0.5970} & \colorPCS{0.3863} & \colorPCS{0.9061} & \colorPCS{0.7829} \\
\hline 
\textbf{GhostSpec-corr} & Data-Free & \colorGhostCorr{0.9992} & \colorGhostCorr{0.7595} & \colorGhostCorr{1.0000} & \colorGhostCorr{1.0000} & \colorGhostCorr{0.8967} & \colorGhostCorr{0.7045} \\
\textbf{GhostSpec-mse} & Data-Free & \colorGhostMSE{0.9760} & \colorGhostMSE{0.9532} & \colorGhostMSE{0.9761} & \colorGhostMSE{0.9761} & \colorGhostMSE{0.9727} & \colorGhostMSE{0.9653} \\
\hline
\multirow{3}{*}{\textbf{Method}} & \multirow{3}{*}{\textbf{Data Dep.}}
& \multicolumn{2}{c|}{\cellcolor{tableheadercolor}\textbf{Structured Pruning~($\uparrow$)}}
& \multicolumn{2}{c|}{\cellcolor{tableheadercolor}\textbf{Merging \& Expansion~($\uparrow$)}}
& \multicolumn{2}{c}{\cellcolor{tableheadercolor}\textbf{Unrelated Models~($\downarrow$)}} \\
\cline{3-8}
& & \begin{tabular}[c]{@{}c@{}}Sheared-Llama\\1.3B\end{tabular}
& \begin{tabular}[c]{@{}c@{}}Sheared-Llama\\2.7B\end{tabular}
& \begin{tabular}[c]{@{}c@{}}Llama2-7b-func\\-call-slerp\end{tabular}
& Camelidae-8x7B
& \begin{tabular}[c]{@{}c@{}}Qwen2.5-7B\end{tabular}
& OPT-6.7b
\\
\hline
\textbf{QueRE} & Data-Aware & \colorQueRE{1.0000} & \colorQueRE{1.0000} & \colorQueRE{0.0910} & \colorQueRE{1.0000} & \colorQueRE{0.3410} & \colorQueRE{1.0000} \\
\textbf{Logits} & Data-Aware & \colorLogits{1.0000} & \colorLogits{1.0000} & \colorLogits{1.0000} & \colorLogits{0.9500} & \colorLogits{0.9967} & \colorLogits{0.2200} \\
\textbf{REEF} & Data-Aware & \colorREEF{0.9315} & \colorREEF{0.9487} & \colorREEF{0.9996} & \colorREEF{0.9991} & \colorREEF{0.2513} & \colorREEF{0.2692} \\
\textbf{PCS} & Data-Free & \colorPCS{0.0000} & \colorPCS{0.0000} & \colorPCS{0.9993} & \colorPCS{0.0204} & \colorPCS{0.0000} & \colorPCS{0.0000} \\
\hline 
\textbf{GhostSpec-corr} & Data-Free & \colorGhostCorr{0.9398} & \colorGhostCorr{0.9414} & \colorGhostCorr{0.9998} & \colorGhostCorr{0.9999} & \colorGhostCorr{0.2940} & \colorGhostCorr{0.3423} \\
\textbf{GhostSpec-mse} & Data-Free & \colorGhostMSE{0.8886} & \colorGhostMSE{0.9045} & \colorGhostMSE{0.9760} & \colorGhostMSE{0.9761} & \colorGhostMSE{0.0000} & \colorGhostMSE{0.5025} \\
\hline\hline
\multicolumn{8}{c}{\cellcolor{primaryheadercolor}\textbf{Primary Model: Mistral-7B}} \\
\hline
\multirow{3}{*}{\textbf{Method}} & \multirow{3}{*}{\textbf{Data Dep.}}
& \cellcolor{tableheadercolor}\textbf{Fine-tuning~($\uparrow$)}
& \cellcolor{tableheadercolor}\textbf{Merging~($\uparrow$)}
& \cellcolor{tableheadercolor}\textbf{Expansion~($\uparrow$)}
& \cellcolor{tableheadercolor}\textbf{Pruning~($\uparrow$)}
& \multicolumn{2}{c}{\cellcolor{tableheadercolor}\textbf{Unrelated Models~($\downarrow$)}} \\
\cline{3-8}
& & \begin{tabular}[c]{@{}c@{}}OpenHermes-2.5\\-Mistral-7B\end{tabular}
& Triunvirato-7b
& \begin{tabular}[c]{@{}c@{}}Chunky-Lemon\\-Cookie-11B\end{tabular}
& \begin{tabular}[c]{@{}c@{}}OpenHermes-2.5\\-Mistral-7B-pruned50\end{tabular}
& Qwen2.5-7B
& Yi-1.5-6B \\
\hline
\textbf{QueRE} & Data-Aware & \colorQueRE{1.0000} & \colorQueRE{1.0000} & \colorQueRE{1.0000} & \colorQueRE{1.0000} & \colorQueRE{0.3410} & \colorQueRE{0.0819} \\
\textbf{Logits} & Data-Aware & \colorLogits{0.9933} & \colorLogits{0.9967} & \colorLogits{1.0000} & \colorLogits{0.9867} & \colorLogits{0.9567} & \colorLogits{0.2067} \\
\textbf{REEF} & Data-Aware & \colorREEF{0.8949} & \colorREEF{0.8538} & \colorREEF{0.8495} & \colorREEF{0.8596} & \colorREEF{0.7473} & \colorREEF{0.8301} \\
\textbf{PCS} & Data-Free & \colorPCS{0.9999} & \colorPCS{0.9997} & \colorPCS{0.8987} & \colorPCS{0.9979} & \colorPCS{0.0000} & \colorPCS{0.0000} \\
\hline 
\textbf{GhostSpec-corr} & Data-Free & \colorGhostCorr{0.9999} & \colorGhostCorr{0.9997} & \colorGhostCorr{0.9981} & \colorGhostCorr{0.9896} & \colorGhostCorr{0.2708} & \colorGhostCorr{0.4304} \\
\textbf{GhostSpec-mse} & Data-Free & \colorGhostMSE{0.9760} & \colorGhostMSE{0.9759} & \colorGhostMSE{0.9758} & \colorGhostMSE{0.9753} & \colorGhostMSE{0.0083} & \colorGhostMSE{0.0581} \\
\hline\hline
\end{tabular}
}
\caption{Comprehensive comparison of fingerprinting methods against various derivative and unrelated models, with \textbf{Llama-2-7b} and \textbf{Mistral-7B} as primary models. The table evaluates robustness to fine-tuning, architectural dissimilarity, compression, merging, expansion, and adversarial transformations. Similarity scores are color-coded based on method-specific thresholds: \colorbox{mygreen}{\phantom{xx}} indicates a score above the threshold (positive classification), while \colorbox{myred}{\phantom{xx}} indicates a score below it (negative classification).}
\label{tab:main_comparison_compact}
\end{table*}

\begin{algorithm}[t]
\caption{Penalty-based Optimal Spectral Alignment}
\label{alg:alignment}
\textbf{Input}: Distance matrix $D \in \mathbb{R}^{N \times M}$ (assume $N \le M$), \\gap penalty $\rho$. \\
\mbox{\textbf{Output}: Average distance along the optimal path.}
\begin{algorithmic}[1]
\STATE Initialize cost matrix $C \in \mathbb{R}^{N \times M}$ and backtrack \\ matrix $B \in \mathbb{Z}^{N \times M}$.
\FOR{$j = 0$ \TO $M-1$}
    \STATE $C[0, j] \gets D[0, j]$ \COMMENT{Base case for the first layer.}
\ENDFOR

\FOR{$i = 1$ \TO $N-1$}
    \FOR{$j = i$ \TO $M-1$}
        \STATE $k^* \gets \arg\min\limits_{i-1 \le k < j} ( C[i-1, k] + (j-k-1)\rho )$
        \STATE $C[i, j] \gets C[i-1, k^*] + (j-k^*-1)\rho + D[i, j]$
        \STATE $B[i, j] \gets k^*$
    \ENDFOR
\ENDFOR

\STATE $j_{\text{end}} \gets \arg\min\limits_{N-1 \le k < M} C[N-1, k]$
\STATE Reconstruct optimal path $P$ by backtracking from \\ $(N-1, j_{\text{end}})$ using $B$.
\STATE \textbf{return} $\frac{1}{|P|}\sum\limits_{(i,j) \in P} D[i,j]$ \COMMENT{Avg. MSE on path.}
\end{algorithmic}
\end{algorithm}




\section{Methodology}
In practice, model weights may undergo transformation attacks such as permutation or scaling, which preserve the model’s functionality while significantly altering the raw weight distribution~\citep{zhang2024reef}. Directly comparing the singular values of individual attention matrices $W_q$, $W_k$, $W_v$, and $W_o$ is therefore highly vulnerable to such obfuscation techniques. To address this, we propose constructing an invariant fingerprint derived from composite matrix products, which are resistant to these transformation attacks. Additionally, we introduce two complementary similarity metrics, GhostSpec-mse and GhostSpec-corr, designed to capture both fine-grained variations and macroscopic structural properties. Together, these components form a robust spectral fingerprinting methodology, as shown in Figure~\ref{fig:main_figure}.

\subsection{The GhostSpec Fingerprint}
To quantify and compare the structural properties of Transformer-based language models, we first define an invariant spectral fingerprint that is robust to permutation and scaling transformations. For a given model $M$ with $L$ layers, we focus on the attention-related weight matrices: $W_q^{(i)}$, $W_k^{(i)}$, $W_v^{(i)}$, and $W_o^{(i)}$ from each layer $i \in \{1, \dots, L\}$.

We define two invariant matrices per layer whose singular value spectra are resilient to functionality-preserving transformations like permutation or scaling:
\begin{equation}
    M_{qk}^{(i)} = W_q^{(i)} (W_k^{(i)})^T \quad \text{and} \quad M_{vo}^{(i)} = W_v^{(i)} W_o^{(i)}.
\end{equation}
 These transformations alter the weight distribution without changing the model's output by modifying the weights in pairs or scaling them uniformly. Since the relative relationships between the weights are preserved, the model's core functionality and input-output behavior remain unchanged.
 
The fingerprint for each layer is composed of the two singular value vectors derived from these products. The layer fingerprint $\mathcal{S}_M^{(i)}$ is defined as the tuple:
\begin{equation}
    \mathcal{S}_M^{(i)} = \left( \mathbf{s}_{qk,M}^{(i)},\; \mathbf{s}_{vo,M}^{(i)} \right),
\end{equation}
where $\mathbf{s}_{p,M}^{(i)} = \mathrm{SVD}(M_{p,M}^{(i)})$ for $p \in \{qk, vo\}$. The complete model fingerprint $\mathcal{F}_M$ is the sequence of these layer fingerprints across all layers.

\subsection{Similarity Metrics}
Given the GhostSpec fingerprints of two models, $A$ (with $N$ layers) and $B$ (with $M$ layers), we introduce two complementary metrics: GhostSpec-mse, which provides a fine-grained, layer-by-layer comparison to measure structural correspondence, and GhostSpec-corr, a lightweight metric that captures the overall trend of spectral properties across layers. Used together, these metrics offer a comprehensive and reliable assessment of model lineage.

\subsubsection{Fine-grained Similarity: GhostSpec-mse.}
This metric performs a detailed, layer-by-layer comparison of the singular value vectors to populate an aggregate distance matrix $D_{\text{avg}} \in \mathbb{R}^{N \times M}$. Each entry $(D_{\text{avg}})_{ij}$ represents the average spectral distance between layer $i$ of model A and layer $j$ of model B, computed as the mean of MSE over the invariant components:
\begin{equation}
(D_{\text{avg}})_{ij} = \frac{1}{2} \sum_{p \in \{qk, vo\}} \frac{1}{r_{p,ij}} \left\| \hat{\mathbf{s}}_{p,A}^{(i)} - \hat{\mathbf{s}}_{p,B}^{(j)} \right\|_2^2
\end{equation}
where:
\begin{itemize}
    \item $p$ is the invariant product type ($qk$ or $vo$).
    \item $r_{p,ij} = \min(\text{eff\_rank}(\mathbf{s}_{p,A}^{(i)}), \text{eff\_rank}(\mathbf{s}_{p,B}^{(j)}))$ is the minimum of the effective ranks of the two singular value vectors being compared.
    \item $\hat{\mathbf{s}}$ denotes a processed singular value vector. The processing involves first truncating the original vector $\mathbf{s}$ to its top $r_{p,ij}$ values, and then applying min-max normalization to scale the result to the $[0, 1]$ range.
\end{itemize}

To handle models with different depths ($N \neq M$), we apply our POSA algorithm, as shown in Algorithm~\ref{alg:alignment}, to find the minimum-cost alignment path through $D_{\text{avg}}$. The raw similarity score, $d_{\text{path}}$, is the average MSE along this optimal path. Finally, we convert this distance into a normalized similarity score using an inverted Sigmoid transformation:
\begin{equation}
    \mathrm{Sim}_{\text{MSE}}(A, B) = 1 - \frac{1}{1 + e^{-k(d_{\text{path}} - \tau)}},
\end{equation}
where $\tau$ is an empirical discrimination threshold and $k$ is a steepness factor. A score approaching 1.0 indicates high similarity.

\subsubsection{Lightweight Similarity: GhostSpec-corr.}
As illustrated in Figure~\ref{fig:layers_trend_norm}, models with shared lineage exhibit highly correlated spectral trends, while unrelated models show divergent patterns. Based on this observation, we propose the GhostSpec-corr metric, which quantifies model similarity by capturing the overall trend of spectral properties.

First, we generate the trend sequences. For each layer $i$ and component $p$, we compute a scalar value, $\mu_{p,M}^{(i)}$, defined as the mean of the top-$K$ normalized singular values. Here, $K$ is dynamically determined based on the effective rank of each singular value spectrum, consistent with the truncation method used in GhostSpec-mse. This process produces two trend sequences for each model: $\boldsymbol{\mu}_{qk,M}$ and $\boldsymbol{\mu}_{vo,M}$.

Second, we align the sequences. Since these sequences may differ in length, direct comparison is not feasible. To address this, we apply a dynamic sequence alignment algorithm (similar to the POSA algorithm introduced previously) to match the sequence lengths. This step generates new sequences of equal length ($\boldsymbol{\mu}'_{qk,A}$, $\boldsymbol{\mu}'_{qk,B}$, etc.).

Finally, we compute the similarity. We concatenate the aligned sequences for each model and calculate the final similarity score using the distance correlation coefficient:
\begin{equation}
\begin{split}
    \mathrm{Sim}_{\text{Corr}}(A, B) = \mathrm{dCor}\big(&[\boldsymbol{\mu}'_{qk,A}; \boldsymbol{\mu}'_{vo,A}], \\
    &[\boldsymbol{\mu}'_{qk,B}; \boldsymbol{\mu}'_{vo,B}]\big).
\end{split}
\end{equation}
This score quantifies the correlation of the models' high-level spectral evolution, providing a computationally efficient indicator of shared lineage.

\begin{figure}[t]
    \centering
    \includegraphics[width=\columnwidth]{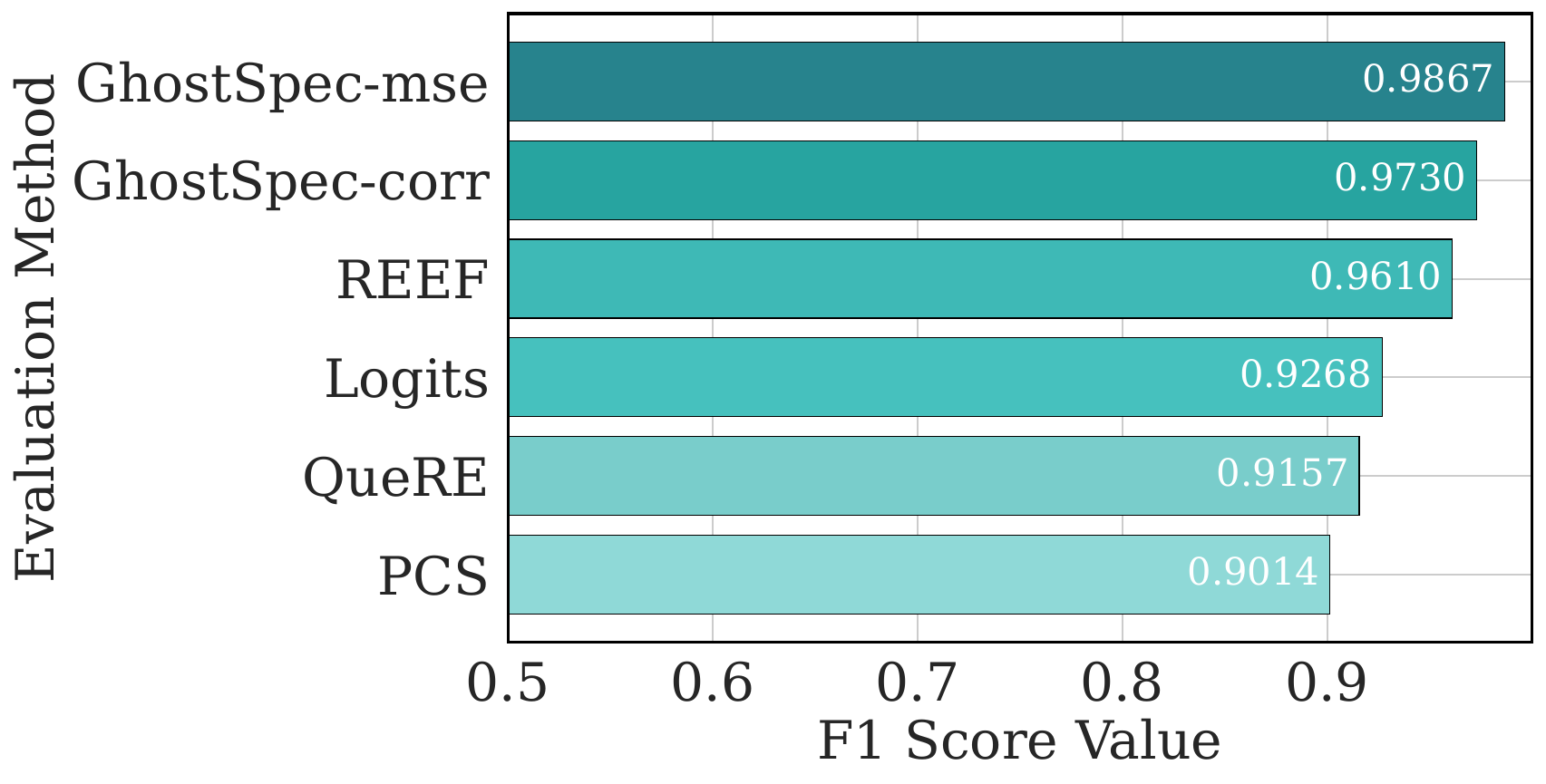}
    \caption{Maximum F1 scores for each method on our dataset. Both GhostSpec variants clearly outperform all baseline methods in accurately distinguishing related from unrelated models.}
    \label{fig:f1_score_comparison}
\end{figure}

\section{Experiments}

To evaluate the effectiveness and robustness of our proposed GhostSpec, we conduct a comprehensive suite of experiments designed to emulate real-world scenarios of model reuse, modification, and transformation.

\subsection{Experimental Setup}
\subsubsection{Dataset Construction.}
We constructed a comprehensive dataset consisting of 55 model pairs, using Llama-2-7b and Mistral-7B as the primary base models. This dataset covers a wide range of transformations, including fine-tuning, compression, merging, and expansion, with ground-truth labels indicating whether each pair is \texttt{related} or \texttt{unrelated}.

\begin{itemize}
    \item \textbf{Fine-Tuning.}
    Fine-tuning is a common source of model derivation. We evaluate similarity between Llama-2-7B and several of its fine-tuned variants: Llama-2-7B-Chat-HF, Vicuna-7B-v1.5 and Llemma-7B, etc. Mistral-7B and several of its fine-tuned variants, including instruction-tuned models, DPO-optimized variants~\citep{nous2024hermes}.

    \item \textbf{Model Pruning.}
    For structured pruning, we consider methods that remove layers based on importance, such as ShortGPT~\citep{men2024shortgpt} and Sheared Llama~\citep{xia2023sheared}. For unstructured pruning, we evaluate representative importance-based methods that prune weights according to their significance, including Wanda~\citep{sun2023simple} and SparseGPT~\citep{frantar2023sparsegpt}. \citet{ma2025model} identifies critical sparsity thresholds near 35\% (structured) and 60\% (unstructured), beyond which model performance rapidly deteriorates. We thus evaluate sparsity levels at 30\%, 50\%, and 70\%.
    
    \item \textbf{Model Merging.}
    Model merging combines weights from multiple source models to create a new model. We evaluate GhostSpec on various merging approaches, including direct weight averaging, methods that reduce parameter interference through pruning and sign conflict resolution~\citep{yadav2023ties}, as well as techniques that merge models by fusing weights or aligning output distributions~\citep{goddard-etal-2024-arcees,wan2024knowledge}.
    
    \item \textbf{Model Upcycling.}
    Model upcycling expands model capacity by introducing new components, such as additional layers or sparsely activated expert modules, including techniques that convert dense LLMs into sparse MoEs through Parameter-Efficient Sparsity Crafting (PESC)~\citep{wu2024parameter}.
    
    \item \textbf{Permutation and Scaling Transformations.}
    To evaluate invariance under adversarial weight manipulations, we apply random permutations to the hidden dimensions of MLP weight matrices. For the attention layers, we apply functionality-preserving similarity transforms, which combine both random rotation and scaling, to the attention projection matrices.
    
    \item \textbf{Unrelated Models.}
    We computed similarity scores between our base models (Llama-2-7b and Mistral-7B) and a diverse set of independently trained LLMs, including pythia-6.9b, opt-6.7b, among others\citep{MosaicML2023Introducing,guo2024deepseek,zhang2022opt}.
\end{itemize}

\subsubsection{Evaluation Protocol.}
To ensure a fair and rigorous comparison with baseline methods, we establish a standardized evaluation protocol. For each method, a model pair is classified as 'related' if its similarity score exceeds a certain threshold. We determine the optimal threshold $\tau^*$ for each method by finding the value that maximizes the F1-score against the ground truth labels (1 for related, 0 for unrelated):
\begin{equation}
\tau^* = \operatorname*{argmax}_{\tau \in \mathcal{S}} \text{F1}(\hat{y}_i(\tau), y_i).
\end{equation}
\noindent where:
\begin{itemize}
    \item $y_i$ is the true label, which is 1 for related pairs (based on ground truth) and 0 for unrelated pairs.
    \item $\hat{y}_i(\tau)$ is the predicted label, which is 1 if the similarity score exceeds the threshold $\tau$, and 0 otherwise.
\end{itemize}
This protocol allows us to evaluate each method at its peak performance, ensuring a fair comparison.

\subsubsection{Baseline Methods.} We compare GhostSpec against representative fingerprinting methods across multiple paradigms.

\begin{itemize}
    \item \textbf{Data-Aware Baselines:} These methods require input data to generate outputs or internal representations for analysis. This category includes black-box approaches like \textbf{QueRE}~\citep{sam2025predicting} and \textbf{Logits}~\citep{yang2024fingerprint}, which analyze model outputs, as well as white-box methods like \textbf{REEF}~\citep{zhang2024reef}, which measures hidden representation similarity. A key characteristic of these methods is their reliance on data, which increases computational cost and necessitates curated datasets.
    
    \item \textbf{Data-Free Baselines:} These methods operate directly on static model weights without requiring any input data. This category includes methods like \textbf{PCS}~\citep{zeng2024huref}, which analyzes invariant submatrices within transformer layers. 

\end{itemize}

\subsection{Main Results}
The overall classification performance of each method, measured by its maximum F1-score on our dataset, is summarized in Figure~\ref{fig:f1_score_comparison}. Our two proposed variants, \textbf{GhostSpec-corr} (F1 = 0.9730) and \textbf{GhostSpec-mse} (F1 = 0.9867), achieve a clear lead, outperforming all data-aware and data-free baselines. 
Table~\ref{tab:main_comparison_compact} presents a selection of illustrative examples from our dataset, showing their detailed similarity scores. These scores are color-coded using the method-specific optimal thresholds, providing an intuitive visualization of the final classification results

GhostSpec demonstrated robust performance across various model transformations. It consistently produces high similarity scores for fine-tuned variants, regardless of training data or objectives. This robustness extends to structural modifications, including aggressive pruning (up to 70\% sparsity), where it accurately recovers source models. It also detects strong similarity in merged and upcycled models, and remains resilient to adversarial transformations like permutation and scaling. Overall, GhostSpec reliably distinguishes true derivatives from unrelated models, showcasing strong discriminative power.

\begin{figure}[t]
\centering
\includegraphics[width=\columnwidth]{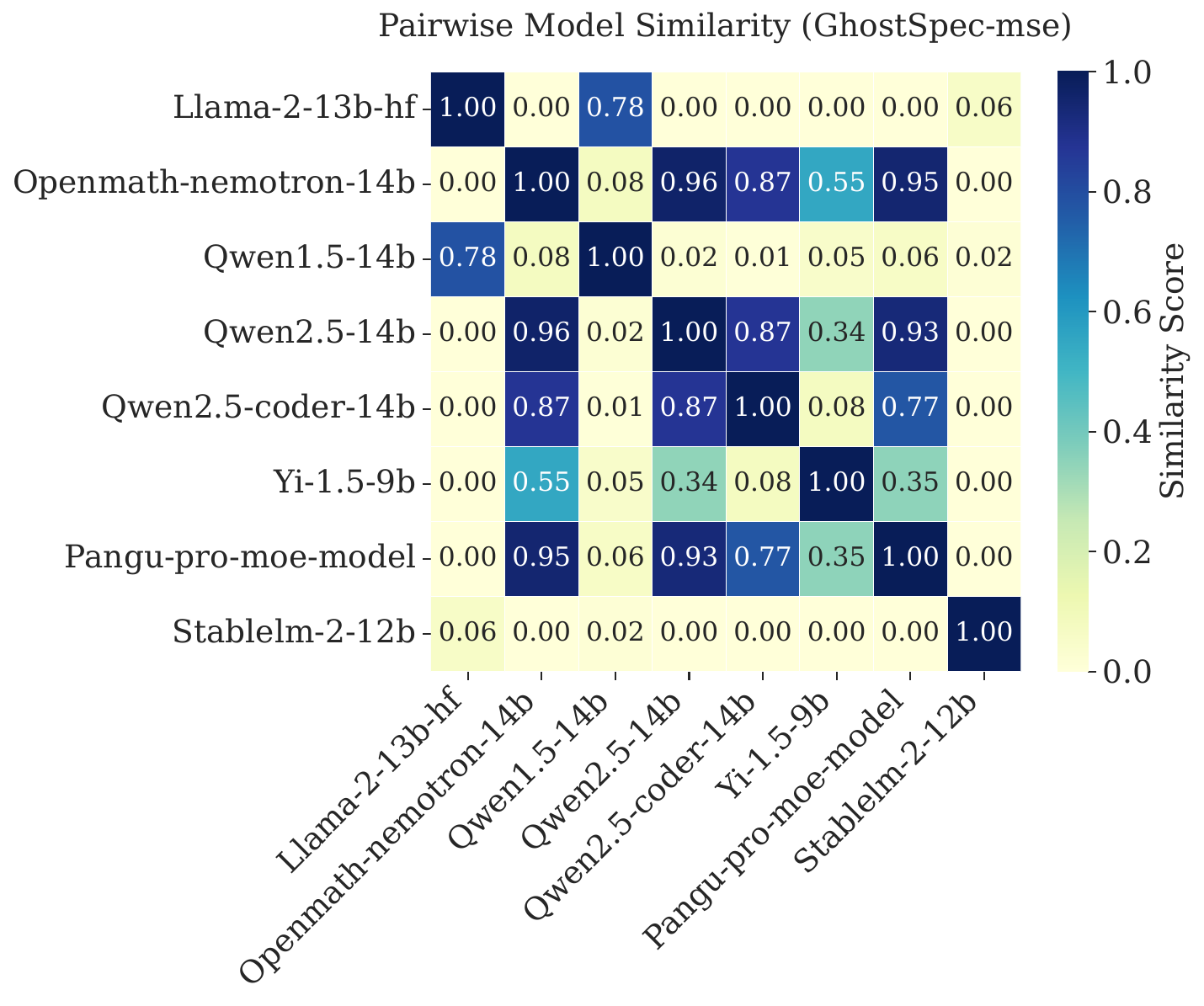}
\caption{Pairwise structural similarity matrix of prominent open-source models computed using GhostSpec-mse. The heatmap visualizes the genealogical relationships between prominent open-source models. Higher scores indicate greater similarity.}
\label{fig:pangu_heatmap_mse}
\end{figure}

\begin{figure}[t]
\centering
\includegraphics[width=\columnwidth]{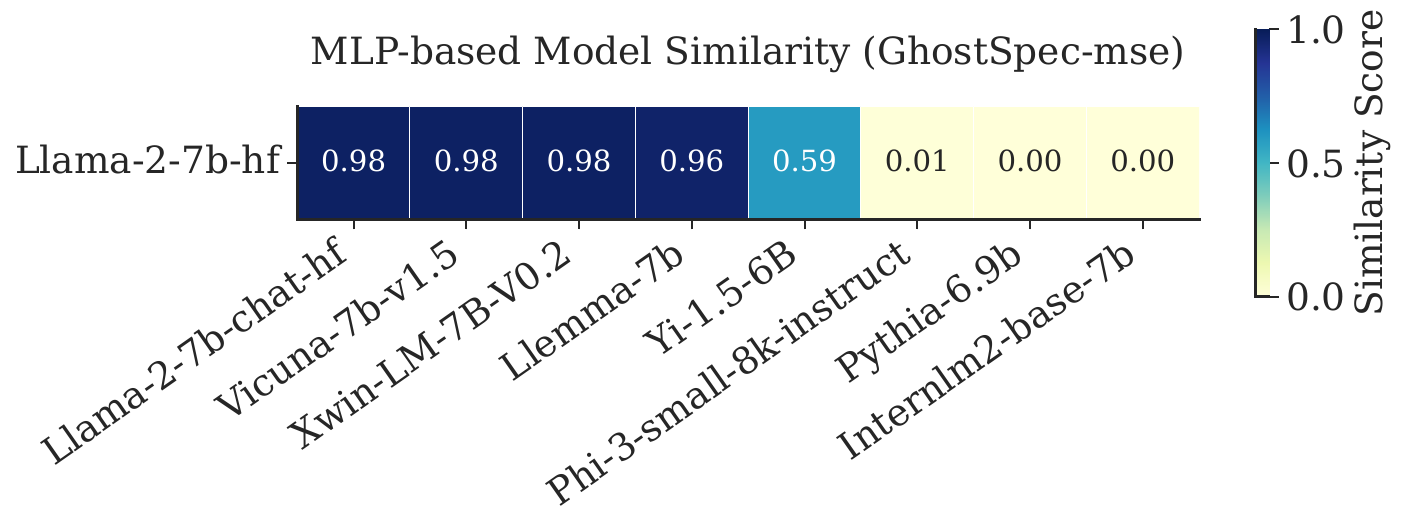}
\caption{MLP-based spectral fingerprint similarity between Llama-2-7b and various models using GhostSpec-mse, computed from singular values of MLP up\_proj and down\_proj weights. Higher scores indicate greater similarity.}
\label{fig:mlp_ghostspec_mse}
\end{figure}

\subsection{Further Discussion}

\subsubsection{Resilience to Evasion Attacks.}

We investigate an evasion strategy wherein an adversary fine-tunes a model with a custom loss to obscure its structural fingerprint. The objective is to maximize spectral distance from a victim model while preserving task performance.

To simulate this, we define a total loss combining task loss and a spectral divergence term:
\begin{equation}
\begin{aligned}
\mathcal{L}_{\text{total}}(\theta) ={} & \mathcal{L}_{\text{task}}(\theta) - \lambda \cdot D_{\text{SVD}}(\mathcal{M}_\theta, \mathcal{M}_{\text{victim}})
\end{aligned}
\end{equation}
where $D_{\text{SVD}}$ denotes the MSE between singular value spectra. The hyperparameter $\lambda$ balances task fidelity and spectral divergence.

We apply this method to adversarially fine-tune a Llama-3.2-1B-instruct model. Results show that it is difficult to significantly alter the singular value spectrum without hurting model performance.

These findings indicate that spectral features are intrinsically tied to model functionality, limiting the practicality of evasion via spectral manipulation.

\subsubsection{Case Study: A High-Profile Lineage Dispute}
We apply \textbf{GhostSpec} using GhostSpec-mse similarity to investigate the recently debated lineage of Pangu-Pro-MoE by comparing it with several open-source models, including the Qwen series and Llama variants.

As shown in Figure~\ref{fig:pangu_heatmap_mse}, GhostSpec finds that Pangu-Pro-MoE has the highest similarity with OpenMath-Nemotron-14B (a fine-tuned variant of Qwen2.5-14B) and Qwen2.5-14B, while showing negligible similarity to unrelated models such as Yi-1.5-9B and Llama-2-13b-hf. 

These results indicate a potential lineage connection between Pangu-Pro-MoE and the Qwen2.5-14B family, though further evidence is needed for confirmation.

\subsubsection{Analysis of MLP Module Spectra.}

To examine whether spectral fingerprinting extends beyond attention layers, we analyze singular values of the MLP up-proj and down-proj weight matrices in each Transformer layer. We compute GhostSpec-mse similarity between Llama-2-7b-hf, its fine-tuned variants, and unrelated models.

As shown in Figure~\ref{fig:mlp_ghostspec_mse}, MLP-based fingerprints effectively distinguish fine-tuned models with similarity above 0.96 from unrelated models showing near-zero similarity.

However, this approach incurs higher computational cost due to the larger size of MLP weight matrices and is less robust to common dense-to-MoE expansions where MLP layers are replaced by experts. Therefore, attention-based fingerprints remain more efficient and structurally stable for reliable lineage verification.

\section{Conclusion}
This paper tackles LLM lineage verification amid widespread reuse and potential plagiarism. We propose GhostSpec, a robust, data-independent, and efficient white-box method that constructs stable fingerprints from the singular value spectra of invariant attention matrix products. Resilient to common modifications and adversarial attacks, it employs a penalty-based optimal path alignment algorithm to handle architectural differences. Extensive experiments demonstrate GhostSpec reliably identifies model ancestry across diverse transformations, including fine-tuning, pruning, merging, and expansion. GhostSpec offers a practical, trustworthy tool to protect intellectual property and improve transparency in open-source AI ecosystems.

\section*{Acknowledgements}
This work was supported by the National Natural Science Foundation of China under Grant No. 62306216, the Fundamental Research Funds for the Central Universities under Grant No. 2042025kf0026, and the Technology Innovation Program of Hubei Province under Grant No. 2024BAB043.
\bibliography{aaai}

\input{dependent_appendix.tex}

\end{document}

%% file: dependent_appendix.tex

\appendix
\section{Theoretical Foundation of the GhostSpec Fingerprint}
\label{app:invariance}

We first characterize the class of potential functionality-preserving weight transformation attacks that can significantly alter model parameters without changing the input-output mapping. To counter these attacks, we identify specific matrix products that remain strictly unaffected. We then provide a formal proof of this invariance, taking into account the practical convention of storing transposed weight matrices.

Throughout this section, we denote the conceptual weight matrices as $W_q, W_k, W_v, W_o$. For computational efficiency (e.g., using row-major input vectors $x$), models typically store their transposed counterparts: $W_q^\top, W_k^\top, W_v^\top, W_o^\top$. A projection is thus calculated as $x \cdot W^\top$. Our derivations will explicitly use this convention.

\subsection{Functionality-Preserving Transformation Attack}

\paragraph{V-O Per-Head Transformation.}
The context vector $c$ from a single attention head is computed by projecting the value vectors $v$ with the output weight matrix $W_o$. The value vectors themselves are projections of the input $x$, i.e., $v = x \cdot W_v^\top$. Thus, the final output $y$ of the head is:
\[
y = (\text{AttentionScores} \cdot (x \cdot W_v^\top)) \cdot W_o^\top.
\]

The transformation involves a block-diagonal invertible matrix $C$ applied to the conceptual weights as:
\[
\tilde{W}_v = C W_v, \quad \tilde{W}_o = W_o C^{-1}.
\]

The corresponding stored weights are:
\[
\tilde{W}_v^\top = W_v^\top C^\top, \quad \tilde{W}_o^\top = (C^{-1})^\top W_o^\top.
\]

The output becomes:
\begin{align*}
\tilde{y} &= (\text{AttentionScores} \cdot (x \cdot \tilde{W}_v^\top)) \cdot \tilde{W}_o^\top \\
    &= (\text{AttentionScores} \cdot (x \cdot W_v^\top C^\top)) \cdot ((C^{-1})^\top W_o^\top) \\
    &= (\text{AttentionScores} \cdot x \cdot W_v^\top) \cdot (C^\top (C^{-1})^\top) \cdot W_o^\top \\
    &= (\text{AttentionScores} \cdot x \cdot W_v^\top) \cdot I^\top \cdot W_o^\top = y.
\end{align*}

\paragraph{Q-K RoPE-Compatible Transformation.}

Self-attention output is computed as:
\begin{equation}
\text{Attention}(Q, K, V) = \text{softmax}\left(\frac{Q K^\top}{\sqrt{d_k}}\right) V,
\end{equation}
where $Q = x W_q^\top$ and $K = x W_k^\top$.

To construct a transformation that preserves this output while remaining compatible with RoPE, we introduce a doubly block-diagonal matrix $P$ and define:
\begin{equation}
\tilde{W}_q = P W_q, \quad \tilde{W}_k = (P^\top)^{-1} W_k.
\end{equation}

The corresponding transformed query and key matrices become:
\begin{equation}
\tilde{Q} = Q P^\top, \quad \tilde{K} = K P^{-1}.
\end{equation}

Then,
\begin{align}
\tilde{Q} \tilde{K}^\top &= (Q P^\top)(K P^{-1})^\top = Q P^\top (P^{-1})^\top K^\top \\
&= Q (P^{-1} P)^\top K^\top = Q K^\top.
\end{align}

Thus, attention logits and softmax remain unchanged. Since $V$ is untouched, the overall attention output is exactly preserved:
\begin{equation}
\text{Attention}(\tilde{Q}, \tilde{K}, V) = \text{Attention}(Q, K, V).
\end{equation}

This proves the transformation modifies $W_q$ and $W_k$ without altering the functionality of the attention layer.

\subsection{Formal Proof of Invariance}

This section provides the formal proof that our spectral fingerprints remain invariant when subjected to the functionality-preserving transformation attacks defined above. This resistance is a cornerstone of the method's robustness. We demonstrate that the singular value spectra of our selected matrix products are strictly immune to these adversarial manipulations.

\paragraph{Invariance of the V-O Product ($M_{ov} = W_o W_v$).}
The transformation is defined on the conceptual weights as $\tilde{W}_v = C W_v$ and $\tilde{W}_o = W_o C^{-1}$. The transformed product is:
\begin{align*}
\tilde{M}_{ov} &= \tilde{W}_o \cdot \tilde{W}_v \\
       &= (W_o C^{-1}) \cdot (C W_v) \\
       &= W_o \cdot (C^{-1} C) \cdot W_v \\
       &= W_o \cdot I \cdot W_v = W_o W_v = M_{ov}.
\end{align*}

The product matrix $M_{ov}$ is strictly invariant. Consequently, its Singular Value Decomposition (SVD), and therefore its singular value spectrum, is also invariant, regardless of the storage convention of the weights.

\paragraph{Invariance of the Q-K Product ($M_{q^\top k} = W_q^\top W_k$).}
The transformation is $\tilde{W}_q = P W_q$ and $\tilde{W}_k = (P^\top)^{-1} W_k$. The transformed product is:
\begin{align*}
\tilde{M}_{q^\top k} &= \tilde{W}_q^\top \cdot \tilde{W}_k \\
          &= (W_q^\top P^\top) \cdot (P^\top)^{-1} W_k \\
          &= W_q^\top \cdot I \cdot W_k = W_q^\top W_k = M_{q^\top k}.
\end{align*}

\section{Methodology and Implementation Details}

This appendix provides specific implementation details for the algorithms and metrics described in Section 4, as well as the hyperparameter values used in our experiments (Section 5), to ensure full reproducibility.

\subsection{Core Function Definitions}

\paragraph{Effective Rank.}
The effective rank, used in Section 4.2 to determine the number of singular values to compare, is a continuous measure of a matrix’s intrinsic dimensionality, defined via the entropy of its normalized singular values. Given a singular value vector $\mathbf{s} = (s_1, s_2, \dots, s_n)$, we define its effective rank as:
\begin{equation}
\text{eff\_rank}(\mathbf{s}) = \exp\left( - \sum_{j=1}^{n} \frac{s_j}{\sum_{i=1}^{n} s_i} \log \frac{s_j}{\sum_{i=1}^{n} s_i} \right)
\end{equation}

\paragraph{Distance Correlation (dCor).}
Distance correlation is used to compute the GhostSpec-corr similarity score (Section 4.2, Equation 7). It is a measure of statistical dependence between two random vectors of arbitrary, not necessarily equal, dimension. Unlike Pearson correlation, distance correlation is zero if and only if the vectors are statistically independent, making it a robust metric for comparing the complex, non-linear trends of our layer-wise spectral mean sequences.

Formally, for two sequences $\mathbf{X} \in \mathbb{R}^{n \times d_1}$ and $\mathbf{Y} \in \mathbb{R}^{n \times d_2}$ with $n$ samples, the empirical distance correlation is computed as follows. First, the Euclidean distance matrices $a, b \in \mathbb{R}^{n \times n}$ are constructed, where $a_{ij} = \|x_i - x_j\|$ and $b_{ij} = \|y_i - y_j\|$. These matrices are then doubly centered:
\begin{equation}
  A_{ij} = a_{ij} - \bar{a}_{i\cdot} - \bar{a}_{\cdot j} + \bar{a}_{\cdot\cdot},
\end{equation}
where $\bar{a}_{i\cdot}$ is the row mean, $\bar{a}_{\cdot j}$ is the column mean, and $\bar{a}_{\cdot\cdot}$ is the grand mean of matrix $a$. The same transformation is applied to $b$ to obtain $B$. The sample distance correlation $\mathcal{R}_n(X, Y)$ is then defined based on the sample distance covariance $\mathcal{V}_n(X, Y)$ and variances $\mathcal{V}_n(X), \mathcal{V}_n(Y)$.

\subsection{Hyperparameter Settings}

\paragraph{POSA Gap Penalty ($\rho$).}
The gap penalty $\rho$ for the Penalty-based Optimal Spectral Alignment (POSA) algorithm (Algorithm 1), which is central to comparing models of varying depth, was set empirically. We observed that a value corresponding to the average single-layer MSE distance between moderately similar models provided a good balance between matching corresponding layers and allowing for necessary skips in pruned or expanded models. Across all experiments, a fixed value of $\rho = 0.002$ was used.

\paragraph{Sigmoid Transformation Parameters ($\tau, k$).}
The parameters for the final inverted Sigmoid transformation (Equation 6), used to convert the raw path distance from POSA into the final GhostSpec-mse similarity score, were determined empirically. The goal was to select values that best separate the distance distributions of related and unrelated model pairs, thereby maximizing the F1-score on a validation set. Based on this analysis, we use a fixed threshold $\tau = 0.00371$ and a steepness factor $k = 1000$.

\paragraph{Method-Specific Thresholds ($\tau^*$).}
As described in the Evaluation Protocol (Section 5.1), we determine an optimal classification threshold $\tau^*$ for each fingerprinting method to ensure a fair and rigorous comparison. This threshold is found by searching for the value that maximizes the F1-score for each method on our ground-truth dataset, ensuring that every method is evaluated at its peak performance. The final, empirically determined thresholds used for all experiments are presented in Table~\ref{tab:optimal_thresholds}.

\begin{table}[h]
\centering
\caption{Optimal classification thresholds for all methods.}
\label{tab:optimal_thresholds}
\begin{tabular}{lc}
\toprule
\textbf{Method} & \textbf{Optimal Threshold ($\tau^*$)} \\
\midrule
QueRE & 0.18 \\
Logits & 0.83 \\
REEF & 0.83 \\
PCS & 0.12 \\
GhostSpec-corr & 0.61 \\
GhostSpec-mse & 0.85 \\
\bottomrule
\end{tabular}
\end{table}

\section{Detailed Experimental Setup}
\label{app:exp_setup}
This appendix provides supplementary details for the experimental setup described in Section 5.

\subsection{Model Dataset}
Our comprehensive dataset is constructed using Llama-2-7B and Mistral-7B as the primary base models. This dataset covers a wide range of transformations, including fine-tuning, compression, merging, and expansion, with ground-truth labels indicating whether each pair is related or unrelated. A complete list of all model pairs and their corresponding modification type is provided in Table 2.

\subsection{Computing Infrastructure}
All experiments were conducted on a server equipped with four NVIDIA GeForce RTX 4090 GPUs (24GB VRAM). The software environment was based on CUDA Version 12.2, with Python 3.10 and PyTorch 2.1.

\subsection{Computational Efficiency Analysis}
\label{app:efficiency}

We analyze the computational efficiency of GhostSpec in two distinct stages: the one-time fingerprint extraction and the pairwise similarity comparison.

\paragraph{Theoretical Complexity.}
\begin{itemize}
    \item \textbf{Stage 1: Fingerprint Extraction.} For a model with $L$ layers and hidden dimension $d$, we compute two invariant products per layer ($M_{qk}$ and $M_{vo}$) and perform Singular Value Decomposition (SVD). While the worst-case complexity for full SVD is $O(d^3)$, our practical implementation employs randomized truncated SVD to extract only the top-$K$ singular values, where $K \ll d$ (typically determined by effective rank). This reduces the practical complexity to $O(L \cdot d^2 K)$. Importantly, this is a one-time cost per model checkpoint.
    
    \item \textbf{Stage 2: Similarity Comparison.} To compare two models with $N$ and $M$ layers, the POSA algorithm computes a pairwise distance matrix and executes dynamic programming to find the optimal alignment. The complexity is $O(N \cdot M \cdot K)$. Crucially, this step is independent of the model's large hidden dimension $d$, making the retrieval and verification process extremely fast.
\end{itemize}

\paragraph{Empirical Runtime.}
We measured execution times on a server equipped with a single NVIDIA GeForce RTX 4090 GPU (24GB VRAM) for extraction and standard Intel Xeon CPUs for comparison. Table~\ref{tab:runtime_extraction} details the fingerprint extraction times for models of varying sizes.

\begin{table}[h]
    \centering
    \small
    \begin{tabular}{lcr}
    \toprule
    \textbf{Model} & \textbf{Parameters} & \textbf{Extraction Time (s)} \\
    \midrule
    Qwen3-1.7B & 1.7B & 30.7 \\
    Qwen3-4B & 4B & 46.6 \\
    Qwen3-8B & 8B & 132.0 \\
    \bottomrule
    \end{tabular}
    \caption{One-time fingerprint extraction times measured on a single RTX 4090 GPU. The process is efficient, taking approximately 2 minutes for an 8B parameter model.}
    \label{tab:runtime_extraction}
\end{table}

The pairwise comparison stage is near-instantaneous. For example, computing the similarity between Llama-2-7B and Qwen2.5-7B takes approximately 0.24 seconds, and comparing Llama-2-7B with the significantly deeper Qwen2.5-32B takes only 0.62 seconds. This efficiency demonstrates that GhostSpec is highly scalable for checking a query model against a large repository of existing fingerprints.

\section{Additional Experiments and Visualizations}
\label{app:ablation}
This section provides additional experimental results that empirically justify key design choices of the GhostSpec method (Section 4) and offers further visual evidence of its effectiveness.

\subsection{Out-of-Sample Threshold Generalization}
\label{app:generalization}

To evaluate the generalizability of our method and the robustness of the optimal thresholds derived in Section 5.1 ($\tau_{\text{mse}} = 0.85$ and $\tau_{\text{corr}} = 0.61$), we conducted out-of-sample evaluations on two additional model families with significantly different parameter scales: the Qwen2.5-3B family and the Qwen2.5-32B family.

We applied the fixed thresholds determined solely from the initial Llama-2/Mistral-7B dataset.

\paragraph{Results on 3B Scale.}
Table~\ref{tab:qwen3b_results} presents the results for the 3B parameter scale. GhostSpec correctly identifies all fine-tuned variants as related, with scores well above the thresholds. Conversely, unrelated models like MiniMA-3B and bloom-3b are correctly rejected.

\begin{table}[h]
    \centering
    \small
    \resizebox{\linewidth}{!}{
    \begin{tabular}{llccr}
    \toprule
    \textbf{Base Model} & \textbf{Comparison Model} & \textbf{GhostSpec-mse} & \textbf{GhostSpec-corr} & \textbf{Ground Truth} \\
    \midrule
    Qwen2.5-3B & Qwen2.5-3B-Instruct & \textbf{0.9761} & \textbf{0.9999} & Related \\
    Qwen2.5-3B & Qwen2.5-Coder-3B & \textbf{0.9618} & \textbf{0.9407} & Related \\
    Qwen2.5-3B & SDLM-3B-D4 & \textbf{0.9760} & \textbf{0.9990} & Related \\
    Qwen2.5-3B & calme-3.1-instruct-3b & \textbf{0.9761} & \textbf{0.9999} & Related \\
    \midrule
    Qwen2.5-3B & MiniMA-3B & 0.0000 & 0.6097 & Unrelated \\
    Qwen2.5-3B & bloom-3b & 0.0000 & 0.2765 & Unrelated \\
    \bottomrule
    \end{tabular}
    }
    \caption{Out-of-sample evaluation on the Qwen2.5-3B family using fixed thresholds ($\tau_{\text{mse}} > 0.85$, $\tau_{\text{corr}} > 0.61$). Scores in \textbf{bold} indicate a positive classification (Related) consistent with Ground Truth.}
    \label{tab:qwen3b_results}
\end{table}

\paragraph{Results on 32B Scale.}
Table~\ref{tab:qwen32b_results} extends this validation to the larger 32B parameter scale. Despite the significant increase in model depth and width compared to the 7B base models used for calibration, GhostSpec maintains its discriminative power. All derivative models are identified with high confidence (MSE $> 0.97$), while unrelated large models show negligible similarity.

\begin{table}[h]
    \centering
    \small
    \resizebox{\linewidth}{!}{
    \begin{tabular}{llccr}
    \toprule
    \textbf{Base Model} & \textbf{Comparison Model} & \textbf{GhostSpec-mse} & \textbf{GhostSpec-corr} & \textbf{Ground Truth} \\
    \midrule
    Qwen2.5-32B & BFS-Prover-V2-32B & \textbf{0.9761} & \textbf{0.9999} & Related \\
    Qwen2.5-32B & K2-Think & \textbf{0.9761} & \textbf{0.9996} & Related \\
    Qwen2.5-32B & Qwen2.5-32B-Instruct & \textbf{0.9761} & \textbf{0.9999} & Related \\
    Qwen2.5-32B & SDLM-32B-D4 & \textbf{0.9761} & \textbf{0.9965} & Related \\
    \midrule
    Qwen2.5-32B & Yi-34B & 0.0003 & 0.2999 & Unrelated \\
    Qwen2.5-32B & opus-v1-34B & 0.0003 & 0.2934 & Unrelated \\
    \bottomrule
    \end{tabular}
    }
    \caption{Out-of-sample evaluation on the Qwen2.5-32B family using fixed thresholds ($\tau_{\text{mse}} > 0.85$, $\tau_{\text{corr}} > 0.61$). The method remains robust at larger scales.}
    \label{tab:qwen32b_results}
\end{table}

These results confirm that the structural fingerprints captured by GhostSpec are scale-invariant and that the decision thresholds established on 7B models generalize effectively to both smaller (3B) and larger (32B) architectures without the need for recalibration.

\subsection{Ablation Studies}

\paragraph{On the Necessity of Invariant Products}
\label{app:invariant_ablation}

To demonstrate the critical role of our invariant product construction (Section 4.1), we performed an ablation study. We compared the performance of the full GhostSpec method against a naive variant that computes fingerprints directly from the singular values of the raw attention weights ($W_q$, $W_k$, $W_v$, $W_o$). The specific computational difference between the two methods is as follows:

\begin{itemize}
  \item \textbf{Full GhostSpec (Invariant Method):} This is the standard method proposed in our paper. For each layer $i$, it first constructs two invariant product matrices whose singular value spectra are robust to functionality-preserving transformations:
  \begin{equation}
    M_{\text{ov}}^{(i)} = W_o^{(i)} W_v^{(i)} \quad \text{and} \quad M_{q^\top k}^{(i)} = (W_q^{(i)})^\top W_k^{(i)}
  \end{equation}
  The final fingerprint for the layer, $\mathcal{F}^{(i)}$, is then derived from the singular value vectors of these two product matrices:
  \begin{equation}
    \mathcal{F}^{(i)} = \left( \text{SVD}(M_{\text{ov}}^{(i)}),\; \text{SVD}(M_{q^\top k}^{(i)}) \right)
  \end{equation}

  \item \textbf{Naive Variant (Vulnerable Method):} For the ablation study, we created a naive version that bypasses the invariant product construction. Instead, it directly computes the SVD of the four individual attention weight matrices for each layer $i$:
  \begin{equation}
    W_q^{(i)},\quad W_k^{(i)},\quad W_v^{(i)},\quad W_o^{(i)}
  \end{equation}
  The fingerprint in this case is composed of the singular value vectors from these four raw matrices:
  \begin{equation}
  \begin{split}
    \mathcal{F}_{\text{naive}}^{(i)} = \big( & \text{SVD}(W_q^{(i)}), \text{SVD}(W_k^{(i)}), \\
    & \text{SVD}(W_v^{(i)}), \text{SVD}(W_o^{(i)}) \big)
  \end{split}
  \end{equation}
\end{itemize}

We tested both methods on the task of identifying a transformed version of Llama-2-7b-chat-hf that had undergone both scaling and permutation attacks. The results are shown in Table~\ref{tab:ablation_invariance}. The naive method fails completely, with similarity scores dropping significantly. In contrast, GhostSpec maintains near-perfect similarity, proving that the invariant product construction is essential for robustness against these attacks.

\begin{table}[h]
\centering
\caption{Comparison between the naive direct SVD method and the invariant GhostSpec method when faced with an adversarial transformation. Scores are similarity to the original model.}
\label{tab:ablation_invariance}
\scalebox{0.9}{
\begin{tabular}{lcc}
\hline\hline
\textbf{Method} & \textbf{GhostSpec-mse} & \textbf{GhostSpec-corr} \\
\hline
Direct SVD (Vulnerable) & 0.2585 & 0.5684 \\
\textbf{GhostSpec (Invariant)} & \textbf{0.9761} & \textbf{1.0000} \\
\hline\hline
\end{tabular}
}
\end{table}

\begin{figure*}[t]
  \centering
  \begin{subfigure}[t]{0.48\textwidth}
    \centering
    \includegraphics[width=\linewidth]{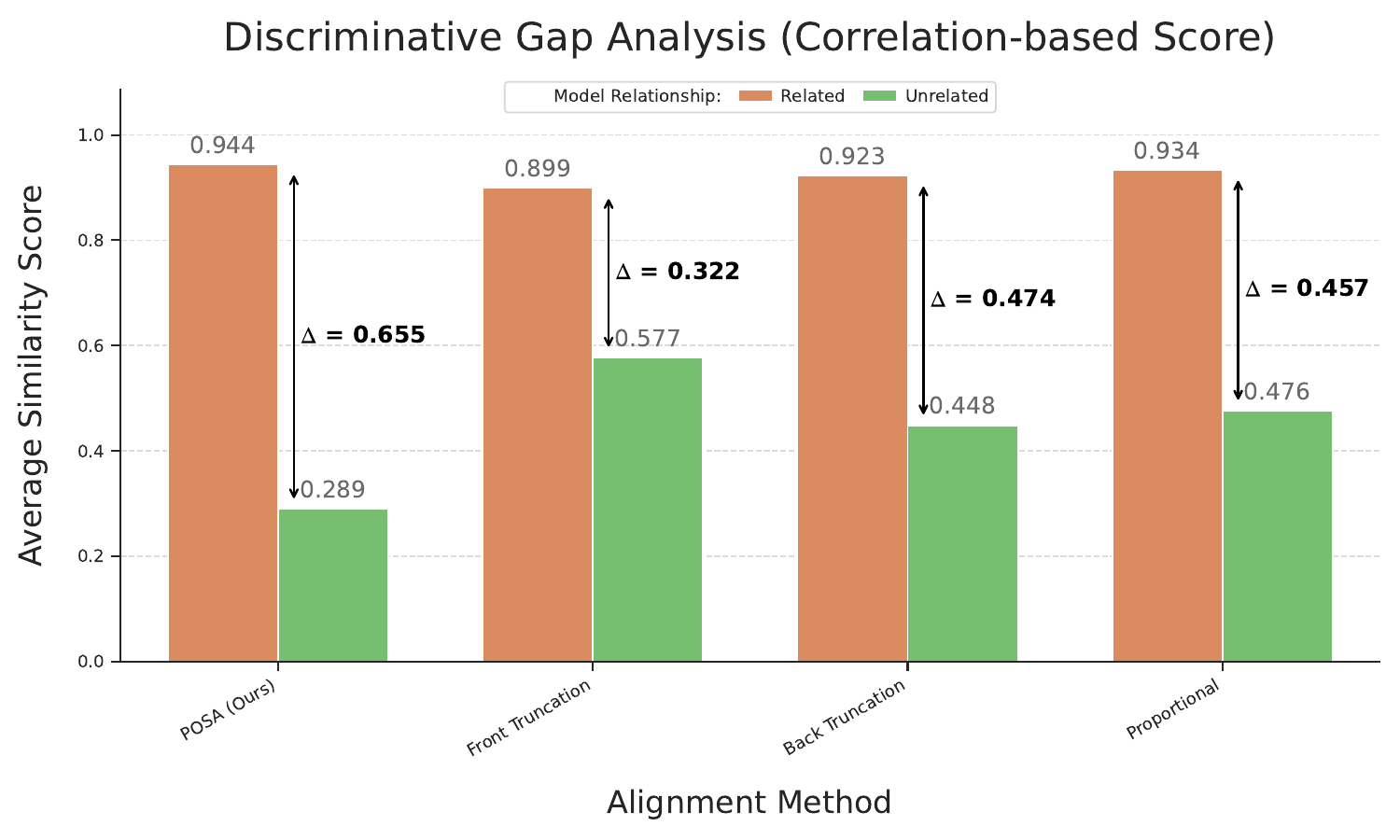}
    \caption{Discriminative gap for Correlation-based scores.}
    \label{fig:posa_ablation_corre}
  \end{subfigure}
  \hfill
  \begin{subfigure}[t]{0.48\textwidth}
    \centering
    \includegraphics[width=\linewidth]{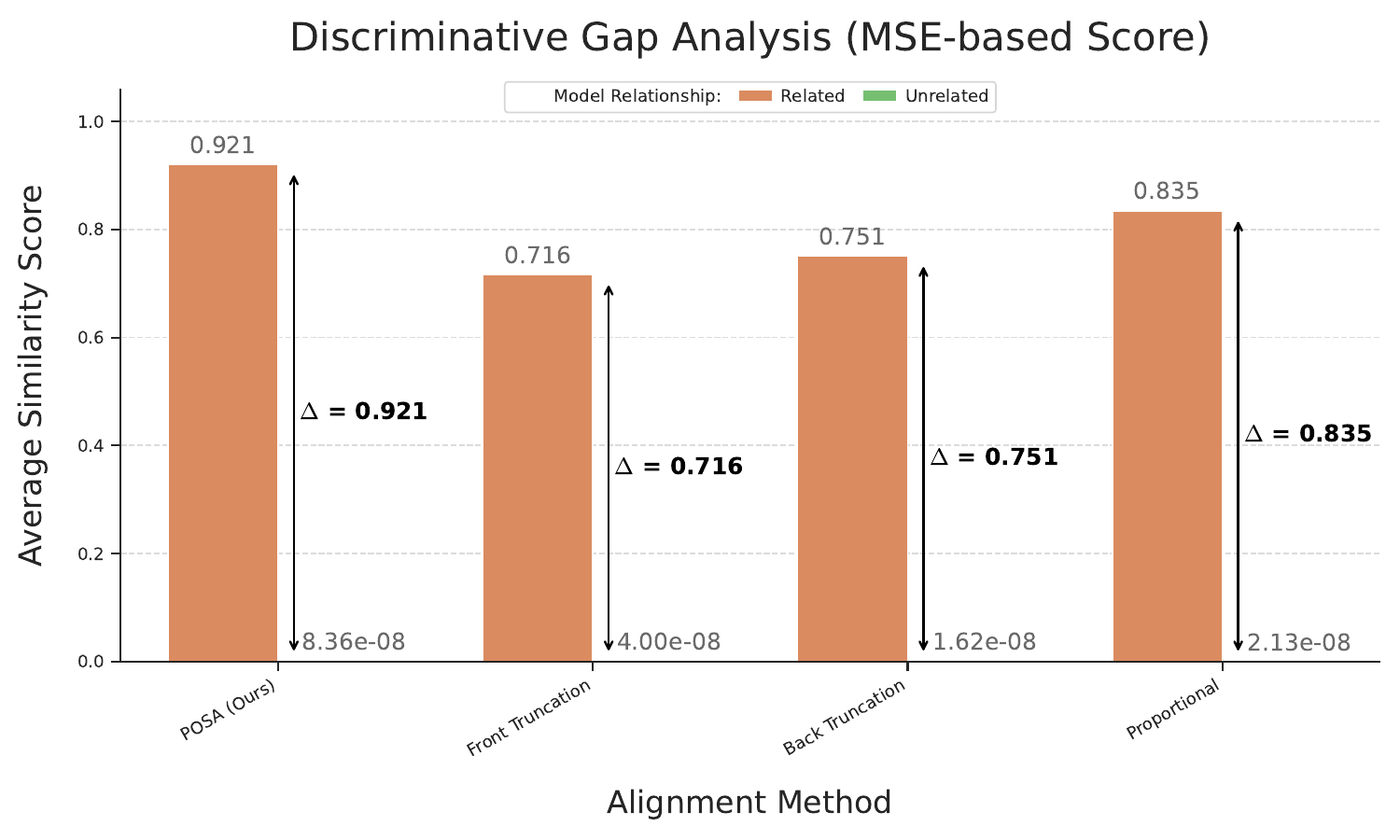}
    \caption{Discriminative gap for MSE-based scores.}
    \label{fig:posa_ablation_mse}
  \end{subfigure}
  \caption{Comparison of the discriminative gaps for two scoring methods. POSA achieves the largest gaps, demonstrating superior discriminative power.}
  \label{fig:posa_ablation_combined}
\end{figure*}

\paragraph{Ablation on Invariant Product Components}
\label{app:product_ablation}

To justify the design choice of using two invariant products ($M_{qk} = W_q W_k^\top$ and $M_{vo} = W_v W_o$) rather than a single one, we performed an ablation study on 28 model pairs (20 related, 8 unrelated). We evaluated the discriminative power of fingerprints derived from:
\begin{enumerate}
    \item \textbf{QK Product Only:} Using only the singular values of $M_{qk}$.
    \item \textbf{VO Product Only:} Using only the singular values of $M_{vo}$.
    \item \textbf{Both (Proposed):} Using the concatenated fingerprints from both products.
\end{enumerate}

As shown in Table~\ref{tab:product_ablation}, while each product independently captures meaningful structural information, combining them yields the highest discriminative gap for both metrics. This confirms that the $Q$-$K$ and $V$-$O$ interactions encode complementary information about the attention mechanism's identity, and using both provides the most robust fingerprint.

\begin{table}[h]
    \centering
    \small
    \begin{tabular}{lcc}
    \toprule
    \textbf{Fingerprint Configuration} & \textbf{$\Delta_{\text{mse}}$} & \textbf{$\Delta_{\text{corr}}$} \\
    \midrule
    QK Product Only ($M_{qk}$) & 0.7099 & 0.3962 \\
    VO Product Only ($M_{vo}$) & 0.5739 & 0.4568 \\
    \textbf{Both (Proposed)} & \textbf{0.7370} & \textbf{0.4672} \\
    \bottomrule
    \end{tabular}
    \caption{Ablation study on the contribution of invariant matrix products. Combining both QK and VO products consistently yields the highest discriminative gap ($\Delta$).}
    \label{tab:product_ablation}
\end{table}

\paragraph{On the Necessity of the POSA Algorithm.}

To demonstrate the necessity of our \textit{Penalty-based Optimal Spectral Alignment (POSA)} algorithm for comparing models of different depths, we conducted a targeted ablation study. The core challenge when comparing models A and B with different layer counts ($N_A \neq N_B$) is establishing a meaningful correspondence between their layer-wise feature sequences, $\mathcal{F}_A = (f_A^{(1)}, \dots, f_A^{(N_A)})$ and $\mathcal{F}_B = (f_B^{(1)}, \dots, f_B^{(N_B)})$. Our experiment was specifically designed to test this scenario.

We selected Llama-2-7b (32 layers) as the base model and compared it against a curated set of seven models, all of which have different layer counts, making a sophisticated alignment strategy essential. These models were categorized as:
\begin{itemize}
  \item \textbf{Related Models}: Five models derived from Llama-2 but with varying depths, including LLaMA-Pro-8B (40 layers), Llama-2-7b-chat-hf-30-sparsity (23 layers), and the Sheared-LLaMA-1.3B series (24 layers).
  \item \textbf{Unrelated Models}: Two architecturally distinct models, Qwen2-7B (28 layers) and Qwen2.5-7B (28 layers).
\end{itemize}

Our experiment compared the performance of POSA against three naive baseline alignment strategies:
\begin{itemize}
  \item \textbf{POSA (Our Method):} Employs dynamic programming with a gap penalty ($\rho$) to find the optimal, potentially non-contiguous, layer alignment that minimizes global distance, as defined by the cost function:
  \[
    \text{Cost}(P) = \sum_{(i_k, j_k) \in P} D_{i_k, j_k} + \sum_{k>1} \rho \cdot (|j_k - j_{k-1}| - 1)
  \]
  \item \textbf{Baseline 1: Front Truncation:} Compares only the first $n$ layers of each model, where $n = \min(N_A, N_B)$.
  \item \textbf{Baseline 2: Back Truncation:} Compares only the last $n$ layers of each model.
  \item \textbf{Baseline 3: Proportional Subsampling:} Downsamples the layer sequence of the larger model to match the length of the smaller one.
\end{itemize}

To quantitatively evaluate the discriminative power of each alignment strategy, we aggregated the results. For each method, we calculated the average similarity score for the Related group and the Unrelated group. The key metric for our evaluation is the Discriminative Gap ($\Delta$), which we define as the difference between these two average scores:
\[
  \Delta = \text{AvgScore}(\text{Related}) - \text{AvgScore}(\text{Unrelated})
\]
A larger gap indicates a superior ability to clearly and reliably distinguish genuine derivatives from unrelated models. The results of this analysis are presented in Figure~\ref{fig:posa_ablation_corre} and Figure~\ref{fig:posa_ablation_mse}.

As shown in the figures, our proposed POSA method consistently achieves the largest discriminative gap across both metrics. For the Correlation-based Score (Figure~\ref{fig:posa_ablation_corre}), POSA achieves a gap of $\Delta = 0.655$, significantly outperforming Front Truncation ($\Delta = 0.322$), Back Truncation ($\Delta = 0.474$), and Proportional Subsampling ($\Delta = 0.457$). Similarly, for the MSE-based Score (Figure~\ref{fig:posa_ablation_mse}), POSA's gap of $\Delta = 0.921$ is substantially larger than that of any other baseline.

These results empirically validate the superiority of the POSA algorithm. While naive methods can show some distinction, their separation between related and unrelated models is less pronounced and less reliable. By finding an optimal layer-wise correspondence that accounts for structural modifications, POSA maximizes the similarity signal from true derivatives while robustly rejecting unrelated models, thereby providing the most effective method for verifying model lineage when architectures differ.

\paragraph{Sensitivity Analysis of POSA Penalty ($\rho$)}
\label{app:posa_sensitivity}

The Penalty-based Optimal Spectral Alignment (POSA) algorithm relies on a gap penalty parameter, $\rho$, to balance the cost of matching dissimilar layers against the cost of skipping layers (introducing gaps). To validate our choice of $\rho = 0.002$ and analyze the method's sensitivity to this hyperparameter, we conducted an experiment using the identical curated dataset of 7 model pairs described in the ablation study above. This dataset consists of the base Llama-2-7b model compared against the same group of 5 related variants with differing depths and 2 architecturally distinct unrelated models.

We measured the Discriminative Gap ($\Delta$), defined as the difference between the average similarity score of related models and unrelated models:
\begin{equation}
\Delta = \text{AvgScore}(\text{Related}) - \text{AvgScore}(\text{Unrelated}).
\end{equation}
A higher $\Delta$ indicates better separation capability.

Table~\ref{tab:posa_sensitivity} presents the results for $\rho \in [0.0, 0.01]$. The results reveal a stable high-performance plateau for $\rho \in [0.0, 0.002]$, where the discriminative gap remains maximal. Beyond this range (e.g., $\rho \ge 0.005$), the penalty becomes too severe, forcing suboptimal alignments and reducing the gap, particularly for the MSE-based metric. Our default choice of $\rho = 0.002$ sits at the edge of this stable region, providing an optimal balance between alignment flexibility and structural constraint.

\begin{table}[h]
    \centering
    \small
    \begin{tabular}{lccl}
    \toprule
    \textbf{Gap Penalty ($\rho$)} & \textbf{$\Delta_{\text{mse}}$} & \textbf{$\Delta_{\text{corr}}$} & \textbf{Observation} \\
    \midrule
    0.000 & 0.9206 & 0.6385 & Stable Plateau \\
    0.001 & 0.9194 & 0.6385 & Stable Plateau \\
    \textbf{0.002 (Default)} & \textbf{0.9194} & \textbf{0.6386} & \textbf{Optimal Choice} \\
    0.005 & 0.8447 & 0.6381 & Onset of Decline \\
    0.010 & 0.8074 & 0.6340 & Significant Decline \\
    \bottomrule
    \end{tabular}
    \caption{Sensitivity analysis of the POSA gap penalty $\rho$. The Discriminative Gap ($\Delta$) remains stable for small penalties but degrades as $\rho$ increases, justifying our selection of 0.002.}
    \label{tab:posa_sensitivity}
\end{table}

\subsection{Visualization of the Fingerprint Space}

To visualize the global structure of the fingerprint space and the discriminative power of our method, we perform dimensionality reduction using the t-SNE algorithm. Instead of constructing explicit high-dimensional feature vectors for each model, we leverage the final similarity scores to build a comprehensive pairwise distance matrix, which directly captures the relationships between all models in our dataset. This process is executed as follows:

\begin{enumerate}
\item \textbf{Pairwise Distance Matrix Construction:} For a set of $N$ models in our dataset, we first compute an $N \times N$ similarity matrix $S$. Each element $S_{ij}$ in this matrix represents the similarity score between model $i$ and model $j$, calculated using either the GhostSpec-mse or GhostSpec-corr method. This similarity matrix is then converted into a distance matrix $D$, where each element is defined as:
\[
D_{ij} = 1 - S_{ij}
\]
The matrix $D$, where $D_{ii} = 0$ and $D_{ij} = D_{ji}$, serves as the direct input for the t-SNE algorithm.

\item \textbf{t-SNE Dimensionality Reduction:} We use the t-SNE algorithm with the \texttt{metric='precomputed'} setting, which operates directly on our distance matrix $D$. The algorithm seeks to find a low-dimensional embedding (in our case, 2D points $Y = \{y_1, y_2, \dots, y_N\}$) that preserves the local neighborhood structure of the original data. It achieves this by minimizing the Kullback–Leibler (KL) divergence between the joint probability distribution $P$, derived from the pairwise distances in $D$, and the joint probability distribution $Q$ of the low-dimensional points in $Y$:
\[
\mathrm{KL}(P \| Q) = \sum_{i \ne j} p_{ij} \log \frac{p_{ij}}{q_{ij}}
\]
The resulting 2D coordinates $Y$ are then plotted, with each point representing a model.
\end{enumerate}

\begin{figure}[h]
\centering
\includegraphics[width=\columnwidth]{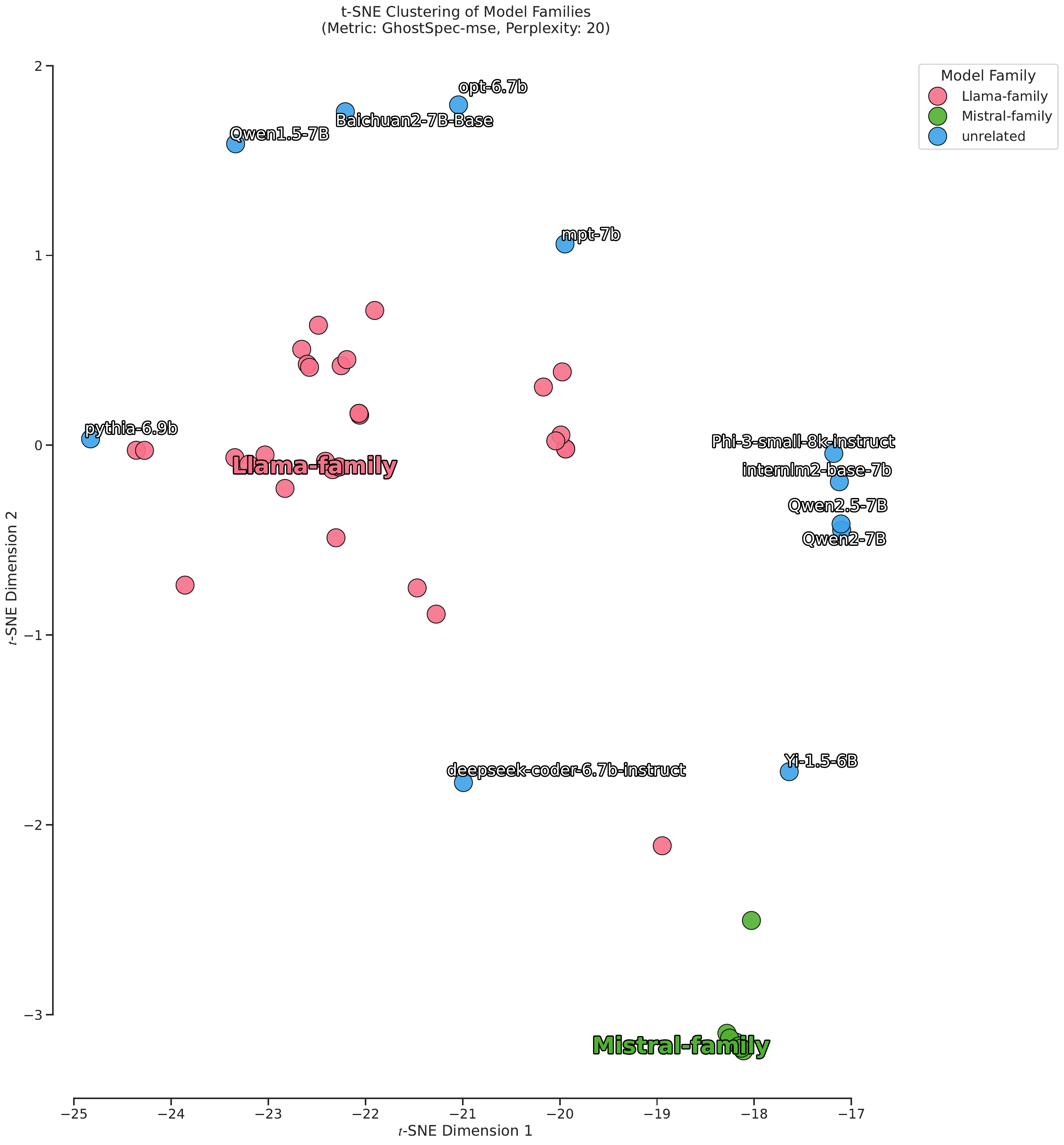}
\caption{t-SNE visualization of the model fingerprint space using the GhostSpec-mse distance metric. Each point represents a model, colored by its family. The clear separation between the Llama family (red), Mistral family (green), and unrelated models (blue) demonstrates that our fingerprint effectively captures genealogical relationships, causing related models to cluster together.}
\label{fig:distance_space_mse_label_path}
\end{figure}

We applied this visualization technique to a diverse dataset comprising models from the Llama family, the Mistral family, and several architecturally distinct unrelated models. The t-SNE embedding based on the GhostSpec-mse metric is shown in Figure~\ref{fig:distance_space_mse_label_path}. The visualization clearly demonstrates the effectiveness of our fingerprinting method. Models belonging to the Llama family form a distinct and tight cluster, while models from the Mistral family converge to a separate region in the embedded space. Critically, the unrelated models remain scattered and do not intrude upon these well-defined family clusters.

This clustering provides strong, intuitive evidence that our GhostSpec similarity scores effectively capture the underlying genealogical relationships between models. The clear separation of model families in the 2D space visually confirms that our method can robustly identify and differentiate model lineages, complementing the quantitative results presented earlier. For reproducibility, the t-SNE parameters were fixed with a perplexity of 20 and a random\_state of 42.

\begin{table*}[h]
\centering
\caption{Complete list of the model pairs used in our dataset for evaluation, supplementing Section 5.1.}
\label{tab:full_model_dataset_appendix}
\begin{tabular}{P{4.5cm} P{3.5cm} | P{4.5cm} P{3.5cm}}
\toprule
\multicolumn{2}{c|}{\textbf{Primary Model: Llama-2-7b}} & \multicolumn{2}{c}{\textbf{Primary Model: Mistral-7B}} \\
\textbf{Comparison Model} & \textbf{Modification Type} & \textbf{Comparison Model} & \textbf{Modification Type} \\
\midrule
Llama-2-7b-chat-hf & Fine-tuning & OpenHermes-2.5-Mistral-7B & Fine-tuning \\
Xwin-LM-7B-V0.2 & Fine-tuning & Nous-Hermes-2-Mistral-7B & Fine-tuning \\
llemma\_7b & Fine-tuning & dolphin-2.2.1-mistral-7b & Fine-tuning \\
CodeLlama-7b-hf & Fine-tuning & & \\
chinese-llama-2-7b & Fine-tuning & & \\
coma-7B-v0.1 & Fine-tuning & & \\
vicuna-7b-v1.5 & Fine-tuning & & \\
\midrule
Sheared-LLaMA-1.3B & Structured Pruning & OpenHermes-pruned50 & Pruning \\
Sheared-LLaMA-2.7B & Structured Pruning & OpenHermes-pruned2.4 & Pruning \\
Sheared-LLaMA-1.3B-Pruned & Structured Pruning & & \\
Sheared-LLaMA-1.3B-ShareGPT & Structured Pruning & & \\
Sheared-LLaMA-2.7B-Pruned & Structured Pruning & & \\
Sheared-LLaMA-2.7B-ShareGPT & Structured Pruning & & \\
Llama-2-chat-shortgpt-30 & Structured Pruning & & \\
\midrule
Llama-2-pruned50-retrained & Unstructured Pruning \\
Llama-2-pruned70-retrained & Unstructured Pruning \\
GBLM-Pruner-LLaMA-2-7B & Unstructured Pruning \\
Llama-2-7b-chat-hf-30-sparsity & Unstructured Pruning \\
Llama-2-7b-gsm8k-pruned\_50 & Unstructured Pruning \\
Llama-2-7b-gsm8k-pruned\_70 & Unstructured Pruning \\
Llama-2-7b-ultrachat200k-pruned\_50 & Unstructured Pruning \\
Llama-2-7b-ultrachat200k-pruned\_70 & Unstructured Pruning \\
\midrule
CodeLlaMa-7B-dare-ties & Merging \& Expansion & Chunky-Lemon-Cookie-11B & Merging \& Expansion \\
llama2-7b-func-calling-slerp & Merging \& Expansion & Triunvirato-7b & Merging \& Expansion \\
FuseLLM-7B & Merging \& Expansion & Daredevil-7B & Merging \& Expansion \\
LLaMA-Pro-8B & Merging \& Expansion & BioMistral-merged-instruct & Merging \& Expansion \\
Camelidae-8x7B & Merging \& Expansion & & \\
Patent-Base-Llama-2-Chat-7B-Slerp & Merging \& Expansion & & \\
\midrule
Llama-2-7b-chat-scaled & Adversarial Transforms \\
Llama-2-7b-chat-permuted & Adversarial Transforms \\
\midrule
Qwen2.5-7B & Unrelated Models & pythia-6.9b & Unrelated Models \\
Yi-1.5-6B & Unrelated Models & mpt-7b & Unrelated Models \\
internlm2-base-7b & Unrelated Models & opt-6.7b & Unrelated Models \\
pythia-6.9b & Unrelated Models & Phi-3-small-8k-instruct & Unrelated Models \\
opt-6.7b & Unrelated Models & Yi-1.5-6B & Unrelated Models \\
Phi-3-small-8k-instruct & Unrelated Models & Qwen2-7B & Unrelated Models \\
 & & Baichuan2-7B-Base & Unrelated Models \\
 &  & Qwen2.5-7B & Unrelated Models \\
 & & deepseek-coder-6.7b & Unrelated Models \\
&  & internlm2-base-7b & Unrelated Models  \\
\bottomrule
\end{tabular}
\end{table*}

%% file: aaai.bib
@article{staats2024small,
  title={Small Singular Values Matter: A Random Matrix Analysis of Transformer Models},
  author={Staats, Max and Thamm, Matthias and Rosenow, Bernd},
  journal={arXiv preprint arXiv:2410.17770},
  year={2024}
}

@article{young2024yi,
  title={Yi: Open foundation models by 01. ai},
  author={Young, Alex and Chen, Bei and Li, Chao and Huang, Chengen and Zhang, Ge and Zhang, Guanwei and Wang, Guoyin and Li, Heng and Zhu, Jiangcheng and Chen, Jianqun and others},
  journal={arXiv preprint arXiv:2403.04652},
  year={2024}
}

@article{cai2024internlm2,
  title={Internlm2 technical report},
  author={Cai, Zheng and Cao, Maosong and Chen, Haojiong and Chen, Kai and Chen, Keyu and Chen, Xin and Chen, Xun and Chen, Zehui and Chen, Zhi and Chu, Pei and others},
  journal={arXiv preprint arXiv:2403.17297},
  year={2024}
}

@inproceedings{biderman2023pythia,
  title={Pythia: A suite for analyzing large language models across training and scaling},
  author={Biderman, Stella and Schoelkopf, Hailey and Anthony, Quentin Gregory and Bradley, Herbie and O’Brien, Kyle and Hallahan, Eric and Khan, Mohammad Aflah and Purohit, Shivanshu and Prashanth, USVSN Sai and Raff, Edward and others},
  booktitle={International Conference on Machine Learning},
  pages={2397--2430},
  year={2023},
  organization={PMLR}
}

@article{DBLP:journals/corr/abs-2310-06825,
  author       = {Albert Q. Jiang and
                  Alexandre Sablayrolles and
                  Arthur Mensch and
                  Chris Bamford and
                  Devendra Singh Chaplot and
                  Diego de Las Casas and
                  Florian Bressand and
                  Gianna Lengyel and
                  Guillaume Lample and
                  Lucile Saulnier and
                  L{\'{e}}lio Renard Lavaud and
                  Marie{-}Anne Lachaux and
                  Pierre Stock and
                  Teven Le Scao and
                  Thibaut Lavril and
                  Thomas Wang and
                  Timoth{\'{e}}e Lacroix and
                  William El Sayed},
  title        = {Mistral 7B},
  journal      = {CoRR},
  volume       = {abs/2310.06825},
  year         = {2023},
  url          = {https://doi.org/10.48550/arXiv.2310.06825},
  doi          = {10.48550/ARXIV.2310.06825},
  eprinttype    = {arXiv},
  eprint       = {2310.06825},
  bibsource    = {dblp computer science bibliography, https://dblp.org}
}

@article{azerbayev2023llemma,
  title={Llemma: An open language model for mathematics},
  author={Azerbayev, Zhangir and Schoelkopf, Hailey and Paster, Keiran and Santos, Marco Dos and McAleer, Stephen and Jiang, Albert Q and Deng, Jia and Biderman, Stella and Welleck, Sean},
  journal={arXiv preprint arXiv:2310.10631},
  year={2023}
}

@article{touvron2023llama,
  title={Llama 2: Open foundation and fine-tuned chat models},
  author={Touvron, Hugo and Martin, Louis and Stone, Kevin and Albert, Peter and Almahairi, Amjad and Babaei, Yasmine and Bashlykov, Nikolay and Batra, Soumya and Bhargava, Prajjwal and Bhosale, Shruti and others},
  journal={arXiv preprint arXiv:2307.09288},
  year={2023}
}

@misc{xwin-lm,
  title = {Xwin-LM},
  author = {{Xwin-LM Team}},
  howpublished = {\url{https://github.com/Xwin-LM/Xwin-LM}},
  year = {2023},
  note = {Accessed: 2023-09}
}

@article{pasquini2024llmmap,
  title={Llmmap: Fingerprinting for large language models},
  author={Pasquini, Dario and Kornaropoulos, Evgenios M and Ateniese, Giuseppe},
  journal={arXiv preprint arXiv:2407.15847},
  year={2024}
}

@article{mcgovern2024your,
  title={Your large language models are leaving fingerprints},
  author={McGovern, Hope and Stureborg, Rickard and Suhara, Yoshi and Alikaniotis, Dimitris},
  journal={arXiv preprint arXiv:2405.14057},
  year={2024}
}

@article{sam2025predicting,
  title={Predicting the performance of black-box llms through self-queries},
  author={Sam, Dylan and Finzi, Marc and Kolter, J Zico},
  journal={arXiv preprint arXiv:2501.01558},
  year={2025}
}

@article{yang2024fingerprint,
  title={A fingerprint for large language models},
  author={Yang, Zhiguang and Wu, Hanzhou},
  journal={arXiv preprint arXiv:2407.01235},
  year={2024}
}

@article{xu2024instructional,
  title={Instructional fingerprinting of large language models},
  author={Xu, Jiashu and Wang, Fei and Ma, Mingyu Derek and Koh, Pang Wei and Xiao, Chaowei and Chen, Muhao},
  journal={arXiv preprint arXiv:2401.12255},
  year={2024}
}

@inproceedings{kirchenbauer2023watermark,
  title={A watermark for large language models},
  author={Kirchenbauer, John and Geiping, Jonas and Wen, Yuxin and Katz, Jonathan and Miers, Ian and Goldstein, Tom},
  booktitle={International Conference on Machine Learning},
  pages={17061--17084},
  year={2023},
  organization={PMLR}
}

@article{nagatsuka2025nested,
  title={A Nested Watermark for Large Language Models},
  author={Nagatsuka, Koichi and Morishita, Terufumi and Sogawa, Yasuhiro},
  journal={arXiv preprint arXiv:2506.17308},
  year={2025}
}

@article{zhang2024reef,
  title={Reef: Representation encoding fingerprints for large language models},
  author={Zhang, Jie and Liu, Dongrui and Qian, Chen and Zhang, Linfeng and Liu, Yong and Qiao, Yu and Shao, Jing},
  journal={arXiv preprint arXiv:2410.14273},
  year={2024}
}

@article{wu2025gradient,
  title={Gradient-Based Model Fingerprinting for LLM Similarity Detection and Family Classification},
  author={Wu, Zehao and Zhao, Yanjie and Wang, Haoyu},
  journal={arXiv preprint arXiv:2506.01631},
  year={2025}
}

@article{liang2025origin,
  title={Origin Tracer: A Method for Detecting LoRA Fine-Tuning Origins in LLMs},
  author={Liang, Hongyu and Zheng, Yuting and Li, Yihan and Zhang, Yiran and Liang, Shiyu},
  journal={arXiv preprint arXiv:2505.19466},
  year={2025}
}

@article{zeng2024huref,
  title={Huref: Human-readable fingerprint for large language models},
  author={Zeng, Boyi and Wang, Lizheng and Hu, Yuncong and Xu, Yi and Zhou, Chenghu and Wang, Xinbing and Yu, Yu and Lin, Zhouhan},
  journal={Advances in Neural Information Processing Systems},
  volume={37},
  pages={126332--126362},
  year={2024}
}

@article{yoon2025intrinsic,
  title={Intrinsic Fingerprint of LLMs: Continue Training is NOT All You Need to Steal A Model!},
  author={Yoon, Do-hyeon and Chun, Minsoo and Allen, Thomas and M{\"u}ller, Hans and Wang, Min and Sharma, Rajesh},
  journal={arXiv preprint arXiv:2507.03014},
  year={2025}
}

@article{workshop2022bloom,
  title={Bloom: A 176b-parameter open-access multilingual language model},
  author={Workshop, BigScience and Scao, Teven Le and Fan, Angela and Akiki, Christopher and Pavlick, Ellie and Ili{\'c}, Suzana and Hesslow, Daniel and Castagn{\'e}, Roman and Luccioni, Alexandra Sasha and Yvon, Fran{\c{c}}ois and others},
  journal={arXiv preprint arXiv:2211.05100},
  year={2022}
}

@article{yang2025qwen3,
  title={Qwen3 technical report},
  author={Yang, An and Li, Anfeng and Yang, Baosong and Zhang, Beichen and Hui, Binyuan and Zheng, Bo and Yu, Bowen and Gao, Chang and Huang, Chengen and Lv, Chenxu and others},
  journal={arXiv preprint arXiv:2505.09388},
  year={2025}
}

@article{achiam2023gpt,
  title={Gpt-4 technical report},
  author={Achiam, Josh and Adler, Steven and Agarwal, Sandhini and Ahmad, Lama and Akkaya, Ilge and Aleman, Florencia Leoni and Almeida, Diogo and Altenschmidt, Janko and Altman, Sam and Anadkat, Shyamal and others},
  journal={arXiv preprint arXiv:2303.08774},
  year={2023}
}

@article{yang2024model,
  title={Model merging in llms, mllms, and beyond: Methods, theories, applications and opportunities},
  author={Yang, Enneng and Shen, Li and Guo, Guibing and Wang, Xingwei and Cao, Xiaochun and Zhang, Jie and Tao, Dacheng},
  journal={arXiv preprint arXiv:2408.07666},
  year={2024}
}

@article{zhu2024survey,
  title={A survey on model compression for large language models},
  author={Zhu, Xunyu and Li, Jian and Liu, Yong and Ma, Can and Wang, Weiping},
  journal={Transactions of the Association for Computational Linguistics},
  volume={12},
  pages={1556--1577},
  year={2024},
  publisher={MIT Press 255 Main Street, 9th Floor, Cambridge, Massachusetts 02142, USA~…}
}

@article{yao2024minicpm,
  title={Minicpm-v: A gpt-4v level mllm on your phone},
  author={Yao, Yuan and Yu, Tianyu and Zhang, Ao and Wang, Chongyi and Cui, Junbo and Zhu, Hongji and Cai, Tianchi and Li, Haoyu and Zhao, Weilin and He, Zhihui and others},
  journal={arXiv preprint arXiv:2408.01800},
  year={2024}
}

@article{ma2025model,
  title={Model Hemorrhage and the Robustness Limits of Large Language Models},
  author={Ma, Ziyang and Li, Zuchao and Zhang, Lefei and Xia, Gui-Song and Du, Bo and Zhang, Liangpei and Tao, Dacheng},
  journa81l={arXiv preprint arXiv:2503.23924},
  year={2025}
}

@article{sun2023deep,
  title={Deep intellectual property protection: A survey},
  author={Sun, Yuchen and Liu, Tianpeng and Hu, Panhe and Liao, Qing and Fu, Shaojing and Yu, Nenghai and Guo, Deke and Liu, Yongxiang and Liu, Li},
  journal={arXiv preprint arXiv:2304.14613},
  year={2023}
}

@inproceedings{tang2025covipal,
  title={CoViPAL: Layer-wise Contextualized Visual Token Pruning for Large Vision-Language Models},
  author={Tang, Zicong and Ma, Ziyang and Wang, Suqing and Li, Zuchao and Zhang, Lefei and Zhao, Hai and Li, Yun and Wang, Qianren},
  booktitle={Findings of the Association for Computational Linguistics: EMNLP 2025},
  pages={20701--20714},
  year={2025}
}

@inproceedings{wang2025parameters,
  title={From Parameters to Performance: A Data-Driven Study on LLM Structure and Development},
  author={Wang, Suqing and Li, Zuchao and Luohe, Shi and Du, Bo and Zhao, Hai and Li, Yun and Wang, Qianren},
  booktitle={Proceedings of the 2025 Conference on Empirical Methods in Natural Language Processing},
  pages={26095--26112},
  year={2025}
}

@inproceedings{yang2025xquant,
  title={XQuant: Achieving Ultra-Low Bit KV Cache Quantization with Cross-Layer Compression},
  author={Yang, Haoqi and Yao, Yao and Li, Zuchao and Qi, Baoyuan and Guoming, Liu and Zhao, Hai},
  booktitle={Proceedings of the 2025 Conference on Empirical Methods in Natural Language Processing},
  pages={9796--9811},
  year={2025}
}

@article{zhang2025segment,
  title={Segment First or Comprehend First? Explore the Limit of Unsupervised Word Segmentation with Large Language Models},
  author={Zhang, Zihong and He, Liqi and Li, Zuchao and Zhang, Lefei and Zhao, Hai and Du, Bo},
  journal={arXiv preprint arXiv:2505.19631},
  year={2025}
}

@inproceedings{hu2025songsong,
  title={SongSong: A Time Phonograph for Chinese SongCi Music from Thousand of Years Away},
  author={Hu, Jiliang and Li, Jiajia and Pan, Ziyi and Chen, Chong and Li, Zuchao and Wang, Ping and Zhang, Lefei},
  booktitle={Proceedings of the AAAI Conference on Artificial Intelligence},
  volume={39(25)},
  pages={26229--26237},
  year={2025}
}

@article{poon2025online,
  title={Online Multi-LLM Selection via Contextual Bandits under Unstructured Context Evolution},
  author={Poon, Manhin and Dai, XiangXiang and Liu, Xutong and Kong, Fang and Lui, John and Zuo, Jinhang},
  journal={arXiv preprint arXiv:2506.17670},
  year={2025}
}

@inproceedings{li2023enhancing,
  title={Enhancing visually-rich document understanding via layout structure modeling},
  author={Li, Qiwei and Li, Zuchao and Cai, Xiantao and Du, Bo and Zhao, Hai},
  booktitle={Proceedings of the 31st ACM international conference on multimedia},
  pages={4513--4523},
  year={2023}
}

@article{men2024shortgpt,
  title={Shortgpt: Layers in large language models are more redundant than you expect},
  author={Men, Xin and Xu, Mingyu and Zhang, Qingyu and Wang, Bingning and Lin, Hongyu and Lu, Yaojie and Han, Xianpei and Chen, Weipeng},
  journal={arXiv preprint arXiv:2403.03853},
  year={2024}
}

@article{xia2023sheared,
  title={Sheared llama: Accelerating language model pre-training via structured pruning},
  author={Xia, Mengzhou and Gao, Tianyu and Zeng, Zhiyuan and Chen, Danqi},
  journal={arXiv preprint arXiv:2310.06694},
  year={2023}
}

@article{sun2023simple,
  title={A simple and effective pruning approach for large language models},
  author={Sun, Mingjie and Liu, Zhuang and Bair, Anna and Kolter, J Zico},
  journal={arXiv preprint arXiv:2306.11695},
  year={2023}
}

@inproceedings{frantar2023sparsegpt,
  title={Sparsegpt: Massive language models can be accurately pruned in one-shot},
  author={Frantar, Elias and Alistarh, Dan},
  booktitle={International conference on machine learning},
  pages={10323--10337},
  year={2023},
  organization={PMLR}
}

@article{yadav2023ties,
  title={Ties-merging: Resolving interference when merging models},
  author={Yadav, Prateek and Tam, Derek and Choshen, Leshem and Raffel, Colin A and Bansal, Mohit},
  journal={Advances in Neural Information Processing Systems},
  volume={36},
  pages={7093--7115},
  year={2023}
}

@inproceedings{goddard-etal-2024-arcees,
    title = "Arcee{'}s {M}erge{K}it: A Toolkit for Merging Large Language Models",
    author = "Goddard, Charles  and
      Siriwardhana, Shamane  and
      Ehghaghi, Malikeh  and
      Meyers, Luke  and
      Karpukhin, Vladimir  and
      Benedict, Brian  and
      McQuade, Mark  and
      Solawetz, Jacob",
    editor = "Dernoncourt, Franck  and
      Preo{\c{t}}iuc-Pietro, Daniel  and
      Shimorina, Anastasia",
    booktitle = "Proceedings of the 2024 Conference on Empirical Methods in Natural Language Processing: Industry Track",
    month = nov,
    year = "2024",
    address = "Miami, Florida, US",
    publisher = "Association for Computational Linguistics",
    url = "https://aclanthology.org/2024.emnlp-industry.36",
    doi = "10.18653/v1/2024.emnlp-industry.36",
    pages = "477--485",
}

@article{wan2024knowledge,
  title={Knowledge fusion of large language models},
  author={Wan, Fanqi and Huang, Xinting and Cai, Deng and Quan, Xiaojun and Bi, Wei and Shi, Shuming},
  journal={arXiv preprint arXiv:2401.10491},
  year={2024}
}

@article{wu2024parameter,
  title={Parameter-Efficient Sparsity Crafting from Dense to Mixture-of-Experts for Instruction Tuning on General Tasks},
  author={Wu, Haoyuan and Zheng, Haisheng and Yu, Bei},
  journal={arXiv preprint arXiv:2401.02731},
  year={2024}
}

@misc{nous2024hermes,
  title = {Nous Hermes 2 Mistral 7B DPO},
  author = {Teknium and {theemozilla} and {karan4d} and {huemin art}},
  year = {2024},
  howpublished = {\url{https://huggingface.co/NousResearch/Nous-Hermes-2-Mistral-7B-DPO}},
}

@article{zhang2022opt,
  title={Opt: Open pre-trained transformer language models},
  author={Zhang, Susan and Roller, Stephen and Goyal, Naman and Artetxe, Mikel and Chen, Moya and Chen, Shuohui and Dewan, Christopher and Diab, Mona and Li, Xian and Lin, Xi Victoria and others},
  journal={arXiv preprint arXiv:2205.01068},
  year={2022}
}

@article{guo2024deepseek,
  title={DeepSeek-Coder: When the Large Language Model Meets Programming--The Rise of Code Intelligence},
  author={Guo, Daya and Zhu, Qihao and Yang, Dejian and Xie, Zhenda and Dong, Kai and Zhang, Wentao and Chen, Guanting and Bi, Xiao and Wu, Yu and Li, YK and others},
  journal={arXiv preprint arXiv:2401.14196},
  year={2024}
}

@misc{MosaicML2023Introducing,
  author = {{MosaicML NLP Team}},
  title = {Introducing MPT-7B: A New Standard for Open-Source, Commercially Usable LLMs},
  year = {2023},
  howpublished = {\url{www.mosaicml.com/blog/mpt-7b}},
  note = {Accessed: 2023-05-05}
}
